\documentclass{aa}  

\usepackage{amssymb}
\usepackage{pifont}
\usepackage{graphicx}
\usepackage{xcolor}
\usepackage{amsmath}
\usepackage{pgfplots}
\usepackage{txfonts}
\usepackage[colorlinks=true,allcolors=blue]{hyperref}

\begin{document}

   \title{Dust attenuation law in JWST galaxies at $z \sim 7-8$}

   \titlerunning{Dust attenuation law in early galaxies}
   \author{V. Markov, S. Gallerani, A. Pallottini, L. Sommovigo, S. Carniani, A. Ferrara, E. Parlanti,  F. Di Mascia
     }
   \authorrunning{Markov et al.}
   \institute{Scuola Normale Superiore, Piazza dei Cavalieri 7, 56126 Pisa, Italy\\
              \email{vladan.markov@sns.it}
             }
   \date{Received XX XX, 2023; accepted XX XX, 2023}

  \abstract
   {Attenuation curves in star-forming galaxies depend on dust chemical composition, content, and grain size distribution. Such parameters in turn are related to intrinsic galaxy properties such as metallicity, star formation rate, and stellar age. Due to the lack of observational constraints at high redshift, dust empirical curves measured in the local Universe (e.g. the Calzetti and the Small Magellanic Cloud (SMC) curves) have been employed to describe the dust attenuation at early epochs.}
   {We exploit the high sensitivity and spectral resolution of the {\it James Webb Space Telescope (JWST)} to constrain the dust attenuation curves in high-$z$ galaxies. Our goals are to: i) check whether dust attenuation curves evolve with redshift, and ii) quantify the dependence of the inferred galaxy properties on the assumed dust attenuation law.}
   {We develop a modified version of the SED fitting code \texttt{BAGPIPES} by including a detailed dust attenuation curve parameterization. Dust parameters are derived, along with galaxy properties, from the fit to the data from (rest-frame) far ultraviolet to millimeter bands.}
   {Once applied to three star-forming galaxies at $z=7-8$, we find that their attenuation curves differ from local templates. One out of three galaxies shows a characteristic $2175 \AA$ bump, typically associated to the presence of small carbonaceous dust grains such as polycyclic aromatic hydrocarbons (PAHs). This is one of the first evidences suggesting the presence of PAHs in early galaxies. Galaxy properties such as stellar mass ($M_*$) and star formation rate (SFR) inferred from SED fitting are strongly affected by the {\it assumed} attenuation curve, though the adopted star formation history also plays a major role (with deviations of up to $\sim 0.4$ dex for these parameters).}
   {Our results highlight the importance of accounting for the potential diversity of dust attenuation laws when analyzing the properties of galaxies at the EoR, whose dust properties are still poorly understood. The application of our method to a larger sample of galaxies observed with {\it JWST} can provide us important insights into the properties of dust and galaxies in the early universe.}
 
   
   \maketitle
%

\section{Introduction}

Most of our understanding of star-forming galaxies relies on the acquisition of photometric and spectroscopic (when available) multi-wavelength observations, ranging from the rest frame ultraviolet (UV) to the far infrared (FIR). Powerful ground-based (e.g. ELT, VLT, ALMA, NOEMA) and spaceborne (e.g. {\it HST}, {\it Spitzer}) telescopes have allowed us to collect data from galaxies up to the Epoch of Reionization (EoR, $z\sim 5-8$; see \citealp{10.1093/mnras/stt702}, \citealp{2016ARA&A..54..761S}, and  \citealp{finkelstein_2016} for reviews), and even at much earlier epochs (up to $z\sim 17$), thanks to the unprecedented sensitivity of {\it JWST} (\citealp{2022ApJ...938L..15C}; \citealp{2023MNRAS.519..157W}; \citealp{2023MNRAS.518.6011D}; \citealp{2023ApJS..265....5H}; \citealp{2023MNRAS.519.1201A}).  

These data carry information about the interaction between the radiation field of stars and the matter enclosed in the interstellar medium (ISM) of galaxies. In this context, dust grains play a key role at any wavelengths: on one hand, dust scatters and absorbs preferentially short-wavelength (mostly optical and UV) photons emitted from stars; on the other hand, dust heated by stellar radiation re-emits the absorbed energy as long-wavelength photons, giving rise to the so-called FIR bump (\citealp{1999ApJ...521...64M}; \citealp{2000ApJ...533..682C}; see \citealp{2003ARA&A..41..241D} for a review). Thus, the interpretation of the spectral energy distribution (SED) of galaxies requires a clear understanding of dust attenuation curves, resulting from the complex interplay between the properties of dust (chemical composition, grains size distribution, mass, temperature) and the relative distribution of dust grains with respect to stars.

Standard dust templates that are broadly adopted for both nearby and high-$z$ galaxies are the "Calzetti" attenuation curve (\citealp{1994ApJ...429..582C, 2000ApJ...533..682C}), derived for the local starburst galaxies, the Milky Way (MW) extinction curve with a characteristic bump at $\sim 2175 \AA$, the steeply rising Small Magellanic Cloud (SMC) extinction curve, and the Large Magellanic Cloud (LMC) curve which is intermediate between the MW and SMC (\citealp{2001ApJ...548..296W}). However, the dust attenuation properties of high-$z$ sources are generally unknown. Moreover, the dust properties are not well described by the conventional dust models, suggesting the evolution of the dust properties through cosmic times (\citealp{2004Natur.431..533M}; \citealp{2007ApJ...661L...9S};  \citealp{2008ApJ...685.1046L}; \citealp{2010A&A...523A..85G}; \citealp{2021MNRAS.506.3946D}; \citealp{2022MNRAS.512...58F}). This is not surprising since dust properties both depend on: 1) the stellar populations responsible for dust formation and their corresponding timescales, 2) on the dust re-processing mechanisms, 3) on the physical conditions of the ISM; which are generally unknown at high redshifts. 

Dust formation requires physical gas conditions of low temperature ($T<2000\ K$) and high density ($n > 10^8 \ cm^{-3}$, see \citealt{2003ssac.proc...37L} and \citealt{2022FrASS...9.8217T}, for reviews), which are expected to characterize both the atmospheres of low and intermediate-mass ($M_* < 8 \ M_{\odot}$) evolved stars during the asymptotic giant branch (AGB) phase (see \citealp{2006A&A...447..553F}; \citealp{2009MNRAS.397.1661V}; \citealp{2013MNRAS.434.2390N}, for reviews) and the expanding ejecta of core-collapse type II supernovae (SNeII, see \citealp{2001MNRAS.325..726T}; \citealp{2007MNRAS.378..973B}, for reviews). The timescales associated to the dust formation in AGB and SNeII are ruled by the lifetimes of these stellar populations ($\sim 10^8-10^9$ and $10^6$ years, respectively; \citealp{2003MNRAS.343..427M}). In the local Universe, dust formation is mostly ascribed to AGB stars, which are more numerous and, at these epochs, have plenty of time to evolve and enrich the ISM of galaxies. The origin of dust at high-$z$ is instead much more uncertain. Star-forming galaxies are now routinely (sparsely) detected at $z > 6$ ($z > 10$), namely when the age of the Universe $t_H < 0.9$ Gyr ($0.5$ Gyr) starts becoming as long as (much shorter then) the AGB star lifetimes. It is thus questionable whether the main stellar sources of dust remain the same for high-$z$ and local star-forming galaxies (\citealp{2009MNRAS.397.1661V}; \citealp{2015MNRAS.451L..70M}; \citealp{2019MNRAS.490..540L}; \citealp{2019A&A...624L..13L}; \citealp{2020A&A...637A..32B}; \citealp{2020A&A...641A.168N}; \citealp{2020MNRAS.497..956S, 2022MNRAS.513.3122S}; \citealp{2022MNRAS.512..989D}). 

Independently from the actual nature of the stellar populations responsible for dust production, the final properties of dust further depend on several, complex mechanisms of dust reprocessing in the ISM which likely modify the original dust properties. Dust grains can be completely destroyed and/or eroded as a consequence of (e.g.) gas-grain sputtering, grain-grain collision (i.e. shattering \citealp{1980ApJ...239..193D}; \citealp{1994ApJ...431..321T}; \citealp{1995ApJ...454..254B}), dust sublimation (\citealp{1993ApJ...402..441L}) occurring in the ISM, SN shocks, and hot plasma. These processes are expected to both decrease the total amount of dust and to shift the grain size distribution towards smaller grains, steepening the attenuation curves. Vice-versa, dust grains can grow into larger grains by accreting gas-phase metals in the dense ISM or in molecular clouds (\citealt{1978ppim.book.....S, 1980ApJ...239..193D, 2009ASPC..414..453D}, but see also \citealt{2016MNRAS.463L.112F}), and/or via coagulation \citep{1993ApJ...407..806C, 2009MNRAS.394.1061H}. In these cases, the overall dust mass content increases and the grain size distribution becomes more populated at larger grains, flattening the attenuation curve. 

These dust re-processing mechanisms are the main constituents of the dust-cycle in galaxies, and are deeply interconnected with the physical properties of galaxies (e.g., gas density, metallicity, star formation, stellar ages, and radiation fields). For instance, higher densities may favor the growth and coagulation into larger grains, while young stellar populations and/or strong radiation fields enhance the efficiency of dust destruction processes. There are several theoretical and observational studies suggesting that the ISM properties of high-$z$ star-forming galaxies differ from their local counterparts (\citealp{2017ApJ...846..105O}; \citealp{2018MNRAS.478.1170C}; \citealp{2018A&A...609A.130L}; \citealp{2019MNRAS.489....1F}; \citealp{2019MNRAS.482.4906P}; \citealp{2019MNRAS.487.1689P}; \citealp{2020MNRAS.495L..22V, 2021MNRAS.505.5543V};  \citealp{2022A&A...663A.172M}). 

For all the aforementioned reasons, there is no physical justification for assuming that the dust attenuation in high-$z$ sources resembles any of the well-studied  dust templates of the local galaxies, such as the Calzetti attenuation curve (e.g., \citealp{2022MNRAS.515.3126I}) or the SMC extinction curve (e.g., \citealp{2022MNRAS.516..975T}). Currently, there is no consensus on the shape of dust attenuation curves in early galaxies, although some attenuation templates seem favored over others (see e.g. \citealp{2018MNRAS.477..552B}, \citealp{2022MNRAS.512...58F}).

In this work, we implement a "Drude-type" parameterization for the wavelength dependence of dust attenuation (\citealp{2008ApJ...685.1046L}) in the Bayesian Analysis of Galaxies for Physical Inference and Parameter EStimation (\texttt{BAGPIPES}) package (\citealp{2018MNRAS.480.4379C, 2019MNRAS.490..417C}), which is a commonly used tool for modelling galaxy spectra and fitting spectroscopic and photometric observations. The goals of the present work are: (i) to characterize high-$z$ galaxies properties without any {\it a-priori} assumption of their unknown attenuation curve; (ii) to infer the dust attenuation curves of a sample of star-forming galaxies at high-$z$.

This paper is organized in the following way: an overview of the NIRSpec {\it JWST} observations is presented in Sect. \ref{Data}, our methodology is outlined in Sect. \ref{method}, we present our results in Sect. \ref{Results}, we discuss our results in Sect. \ref{Discussion}, and finally, we outline our main conclusions in Sect. \ref{Conclusions}.


\section{Spectroscopic observations} \label{Data}

We use publicly available data that are part of the Cosmic Evolution Early Release Science (CEERS) survey (Program ID: 1345, PI: Finkelstein) that targets the Extended Groth Strip (EGS) field. We analyze the CEERS NIRSpec observations taken by $ R \sim 100$ NIRSpec MultiObject Spectroscopy (MOS) which spans over the wavelength range of $\lambda \sim 0.6-5 \ \mu m$  to probe the nebular lines and continuum emission of high-redshift targets. For each target, three adjacent micro shutters on the Micro-Shutter Assembly (MSA) were opened, to create a slitlet in the cross-dispersion axis and a 3-point nod pattern was adopted to achieve a total exposure of 3063 seconds.

We visually inspected all the product level 3 spectra in the Mikulski Archive for Space Telescopes (MAST) archive, and we selected only galaxies without negative features in their spectra, with a bright continuum and nebular lines at a $z >5$. We then downloaded from MAST the level 3 1D spectra for the targets that match our requirements. The list of the three sources used in this work is provided in Table \ref{sources}. The {\it JWST}/NIRSpec spectra of these objects have been analyzed and multiple detections of the rest-frame optical lines (H$\beta$, H$\gamma$, [\ion{O}{II}] $\lambda 3727$, [\ion{O}{III}] $\lambda\lambda 4959, 5007$, and [\ion{Ne}{III}] $\lambda 3869$) have been reported by \cite{2023arXiv230107072T}. Spectroscopic redshifts of these sources have been estimated based on the central wavelengths of the detected lines in the NIRSpec medium resolution grating spectra.  

\begin{table*}[h]
\centering
 \caption{Overview of the high-$z$ galaxies used in this paper (\citealp{2023arXiv230107072T}).
 \label{sources}
 }
\begin{tabular}{lcccccccc}
 \hline \hline
  \noalign{\smallskip}
  ID & R.A. & DEC. & $z_{\rm{sp}}$ & Other ID  \\
   \noalign{\smallskip}
 \hline
 \noalign{\smallskip}
s00717 & 14:20:19.537 & +52:58:19.85 & 6.932 & CEERS-717, J142019.5+525819.9 \\
 \noalign{\smallskip}
s01143 & 14:20:18.482 & +52:58:10.22 & 6.928 & CEERS-1143, J142018.5+525810.2 \\
 \noalign{\smallskip}
s01149 & 14:20:21.531 & +52:57:58.258 &8.175 & CEERS-1149, J142021.5+525758.3  \\ %
 \noalign{\smallskip}
\hline
\end{tabular}
\end{table*}


\section{Method} \label{method}

\subsection{SED fitting procedure}

\texttt{BAGPIPES} is a Python-based tool that is able to generate physically realistic galaxy spectra from FUV to microwave wavelengths, to predict the spectroscopic and photometric observations, and to fit the model to the data, using the MultiNest nested sampling algorithm (\citealp{2008MNRAS.384..449F, 2019OJAp....2E..10F}). \texttt{BAGPIPES} has been widely used by the astronomical community for fitting the {\it JWST} observations of galaxies at the EoR and constraining their properties (\citealp{2022arXiv220711217A}; \citealp{2023MNRAS.518.6011D}; \citealp{2022arXiv220711135L}; \citealp{2022MNRAS.tmpL.128C}; \citealp{2022arXiv220714265T}). 

\texttt{BAGPIPES} constructs the luminosity of a galaxy in the following way (\citealp{2018MNRAS.480.4379C}): 
 \begin{equation}
L_{\rm{\lambda}} = \sum_{j=1}^{N_c}\sum_{i=1}^{N_a} \mathrm{SFR}_{j}(t_i)\mathrm{SSP}(a_{*,i}, \lambda, Z_j)T_{\mathrm{ISM}}^+(a_{*,i}, \lambda)T_{\mathrm{ISM}}^0(a_{*,i}, \lambda)\Delta a_{*,i},
\end{equation}
where Simple Stellar Population ($\rm{SSP}$) models are a function of the wavelength $\lambda$, the age of the stellar population $a_*$\footnote{We make a distinction between times, $t$, measured from the beginning of the Universe, and ages, $a$, measured backwards in time from the redshift $z_{\rm{obs}}$, so that $t_i = t(z_{\rm{obs}}) - a_i$, following \cite{2018MNRAS.480.4379C}.}, metallicity $Z$, and the initial mass function (IMF),  $\mathrm{SFR}_{j}(t_i)$ is the Star Formation History (SFH), $N_c$ and $N_a$ are the number of SFH and stellar age bins, respectively and $\Delta a_{*,i}$ are the widths of age bins. $T_{\mathrm{ISM}}^+(a_{*,i})$ takes into account absorption, line emission, ionized continuum emission, and emission from warm dust present in \ion{H}{II} regions (\citealp{10.1046/j.1365-8711.2001.04260.x}). $T_{\mathrm{ISM}}^0(a_{*,i})$ represents the transmission function of the neutral interstellar medium (ISM), which is caused by both diffuse dust attenuation and emission.

\texttt{BAGPIPES} uses the Stellar Population Synthesis (SPS) models from the 2016 version of the \cite{2003MNRAS.344.1000B} model (\citealp{10.1093/mnras/stw1756}), that is generated by using the Medium-resolution Isaac Newton Telescope library of empirical spectra (MILES; \citealp{2011A&A...532A..95F}) and adopting a \cite{2002MNRAS.336.1188K} IMF. The \texttt{BAGPIPES} code accepts the pre-defined SPS in the shape of grids of SSP models for a wide range of $\lambda$, $a_*$, and $Z$.  

Nebular line and continuum emission are pre-computed with \texttt{CLOUDY}, following the methodology of \cite{2017ApJ...840...44B}. Calculations were performed with version 17.03 of the photoionization code \texttt{CLOUDY}, that is described by \citealp{2017RMxAA..53..385F}. \texttt{CLOUDY} uses grids of SSP models as input spectra and adopts the ionization parameter $U$ as an input that is allowed to vary in the range of $-4 < \log{(U)} < 0$\footnote{The default $-4 < \log{(U)} < -2$ range in the built-in model grids in \texttt{BAGPIPES} has been extended in order to cover a wider range of values measured in galaxies at the EoR.  For instance, \cite{2023arXiv230107072T} measured $-2.6 < \log{(U)} < -1.3$ for a sample of galaxies at $z \sim 7-9$, including the three sources analyzed in this paper.}, by changing the number of hydrogen-ionizing photons $Q_{\rm{H}}$ and assuming a constant hydrogen density of $n = 100 \ \rm{atoms} \ cm^{-3}$. For each value of $\log{(U)}$, $Z$ and $a_*$, \texttt{CLOUDY} provides the output spectrum with contribution from the nebular line and continuum emission. Next, the resulting output spectrum is attenuated by the dust.

In the original implementation, \texttt{BAGPIPES} adopts three fixed models for dust attenuation curves: \cite{2000ApJ...533..682C} model for local starbursts, the \cite{2003ApJ...594..279G} SMC model, the \cite{1989ApJ...345..245C}  MW model; and two more flexible models with free parameters: the modified \cite{2000ApJ...539..718C} model, and the \cite{2018ApJ...859...11S} model. Dust emission is accounted for with a hot dust component which is included in the \texttt{CLOUDY} continuum emission from the ionized (\ion{H}{II}) region and a cold diffuse dust component which is modeled by the gray body emission. 

The modeled galaxy luminosity is redshifted to the observed redshift ($z_{\rm{obs}}$) and transformed into a flux density:
\begin{equation}
f_{\rm{\lambda_{obs}}} = \frac{L_{\rm{\lambda}}}{4\pi D_{L}(z_{\rm{obs}})^2(1+z_{\rm{obs}})} T_{\rm{IGM}}(\lambda, z_{\rm{obs}}), 
\end{equation}
where $D_{L}(z_{\rm{obs}})$ is the luminosity distance as a function of the  redshift of the observation $z_{\rm{obs}}$ and $ T_{\rm{IGM}}\rm{(\lambda, z_{obs})}$ is the transmission function of the InterGalactic Medium (IGM; based on the \citealp{2014MNRAS.442.1805I} model). Besides the redshift $z_{\rm{obs}}$, an additional global parameter of the SED fitting procedure is $\sigma_{\rm{vel}}$, designed to model the effects of dispersion on the observed spectral features.

Galaxy fundamental properties that are not directly constrained, but are derived from the SFH properties are the so-called "living" stellar mass $M_*$, that is, the total stellar mass at the time of observation, the SFR which is averaged over the last 100 Myr, and the mass-weighted age ${\langle a \rangle}_*^{\rm{m}}$, i.e., the age when the stellar mass of galaxies was assembled.
The mass-weighted time ${\langle t \rangle}_*^{\rm{m}}$ is given by:
\begin{equation}
{\langle t \rangle}_*^{\rm{m}} = \frac{\int_0^{t_{\mathrm{obs}}}t\, \mathrm{SFR}(t)\mathrm{d}t}{\int_0^{t_{\rm{obs}}}\mathrm{SFR}(t)\mathrm{d}t} = \frac{\int_0^{t_{\rm{obs}}}t\, \mathrm{SFR}(t)\mathrm{d}t}{M_*^{\rm{form}}},
\end{equation}%
where $M_*^{\rm{form}}$ is  the total stellar mass formed at the time of observation $t_{\rm{obs}} = t(z_{\rm{obs}})$ (e.g., \citealp{2018MNRAS.480.4379C}).   

Once the model spectrum is constructed, we can fit it by using a Bayesian approach by adopting a MultiNest sampling (\citealp{2008MNRAS.384..449F, 2019OJAp....2E..10F}).
We define the prior probability distributions for each of the parameters of a given model. Next, we fit the model to the observed spectrum, and obtain the posterior probability distribution in parameter space. We can constrain the parameters of a given model, by using the best-fit values and $1\sigma$ uncertainties (i.e., the $16^{th}$, $50^{th}$, and $84^{th}$ percentile) of the posterior distribution. Full details on the \texttt{BAGPIPES} SED fitting method are provided in \cite{2018MNRAS.480.4379C}.

\subsection{Star formation history}

In order to explore the dependence of our results on the adopted SFH model, we account for seven different SFH models among the $\sim$ten models currently implemented in \texttt{BAGPIPES}. We consider five parametric SFH models: the constant, double power-law, exponentially declining (exponential hereafter), delayed exponentially declining (delayed hereafter), and lognormal. These  parametric models are frequently adopted in the fitting of the SED of high-$z$ galaxies (\citealp{2018MNRAS.480.4379C, 2019ApJ...873...44C, 2023MNRAS.518L..45C}; \citealp{2021MNRAS.505.3336L}; \citealp{2021MNRAS.501.1568F}). Additionally, we adopt a non-parametric model with a "continuity" and "bursty continuity" prior (\citealp{2019ApJ...876....3L}; \citealp{2022arXiv220605315W}; \citealp{2022ApJ...927..170T}).

As fiducial, we use the non-parametric SFH model with a continuity prior, since with respect to the above mentioned parametric models, it allows us to recover all the standard dust attenuation templates of synthetic galaxy spectra, along with the global physical properties, as discussed in Appendix \ref{fit_syn}.
The non-parametric models allow more flexibility on the shape of the SFH model and generally provide a better fit to the often complex, "true" SFHs of galaxies (\citealp{2019ApJ...876....3L}; \citealp{2022MNRAS.516..975T}; \citealp{2022arXiv220605315W}). Consequently, the non-parametric models are able to recover less biased, fundamental properties of galaxies (\citealp{2019ApJ...876....3L}; \citealp{2020ApJ...904...33L}). 

Compared to the parametric models, where SFR($t$) is modeled as a parametric function of time, the non-parametric models do not assume that SFR can be described as an explicit function of time, which allows more flexibility on the shape of the SFH model. Instead, SFH is modeled by step functions in time, in which SFR is constant within each time bin. We put $N = 7$ time bins, taking into account \cite{2019ApJ...876....3L} (Appendix A) who demonstrated that the results do not vary with the choice of the number of bins as long as $N = 4-14$.

We set the first time bin as $0-10\ \rm{Myr}$ while the rest of the bins are spaced equally in logarithmic lookback time, between 10 Myr and the lookback time at $z = 20$. The parameters of the non-parametric SFH model are the ratios of $\log(\rm{SFR})$ between two consecutive time bins, i.e., $\log(\rm{SFR}_{n}/\rm{SFR}_{n+1})$, where $n$ goes from 1 to $N-1$. Thus, the non-parametric SFH ﬁts  $\log(\rm{SFR}_{n}/\rm{SFR}_{n+1})$, adopting the Student's-$t$ distribution:
\begin{equation}
\mathrm{PDF(x, \nu)} = \frac{\Gamma(\frac{\nu+1}{2})}{\sqrt{\nu\pi}\Gamma(\frac{1}{2})}\left(1+\frac{(x/\sigma)^2}{\nu}\right)^{-\frac{\nu+1}{2}}\, ,
\end{equation}%
where $x = \log(\rm{SFR}_{n}/\rm{SFR}_{n+1})$, $\Gamma$ is the Gamma function, $\sigma$ is a scale factor regulating the width of the distribution, and $\nu = 2$ is the degrees of freedom determining the behavior in the tails of the probability distribution. We fix $\sigma = 0.3$ for the continuity prior and $\sigma = 1$ for the bursty continuity prior.
The Student's-$t$ distribution is chosen for the continuity prior since it explicitly weights against sharp changes in SFR($t$) such as rapid quenching or extreme burst (\citealp{2017ApJ...837..170L}), whereas bursty continuity prior allows a more "bursty" star formation (\citealp{2022ApJ...927..170T}). We direct the reader to \cite{2019ApJ...876....3L} for more details on the non-parametric models with different priors, and the impact of the choice of the adopted prior on the constrained galaxy parameters.

The \texttt{BAGPIPES} code allows us to also consider the following parametric SFH models, where the SFR can be described as a function of time: 
\begin{itemize}
\item constant, $\rm{SFR}(t) = \frac{M_*^{\rm{form}}}{t_{\rm{max}}-t_{\rm{min}}}$, where $t_{\rm{min}}$ and $t_{\rm{max}}$ are the time limits of the entire SFH; 
\item double power-law (DPL), $\rm{SFR}(t) \propto ((t/\tau)^{\alpha} + (t/\tau)^{-\beta})^{-1}$, where $\alpha$ and $\beta$ are the falling and rising slopes, respectively, and $\tau$ is the turnover time; 
\item exponential, $\rm{SFR}(t) \propto e^{-(t-t_*)/\tau_a}$,  where $t_*$ is the time of the SFR peak, and $\tau_a$ is the timescale of declining SFR);
\item delayed, $\rm{SFR}(t) \propto (t-t_*)e^{-(t-t_*)/\tau_a}$;
\item lognormal, $\rm{SFR}(t) \propto \frac{1}{t}e^{-(\ln{t}-t_*)^2/2{\tau_a}^2}$. 
\end{itemize}
 The list of all the free parameters of the parametric SFH models, along with their priors and ranges, is given in Table \ref{params_limits}.

\begin{table}[h]
\centering
 \caption{Free parameters and their priors of the different parametric SFH models.
 \label{params_limits}
 }
\begin{tabular}{lcccccccc}
 \hline \hline
  \noalign{\smallskip}
  Parameter & Limits & Prior & SFH \\
   \noalign{\smallskip}
  \hline
   \noalign{\smallskip}
  $\log{M_*^{\rm{form}}/M_{\odot}}$ & $(7, 12)$ & Uniform & all \\
 \noalign{\smallskip}
$Z/Z_{\odot}$  & $(0.001, 1)$ & Logarithmic & all \\ %
 \noalign{\smallskip}
 $\tau/\rm{Myr}$  & $(1, 1000)$ & Uniform & DPL  \\ %
 \noalign{\smallskip}
 $\alpha$  & $(0.01, 1000)$ & Logarithmic & DPL \\ %
 \noalign{\smallskip}
 $\beta$  & $(0.01, 1000)$ & Logarithmic & DPL \\ %
 \noalign{\smallskip}
 $a_*^{\rm{min}}/\rm{Myr}$  & $(1, 1000)$ & Uniform & Constant \\ %
 \noalign{\smallskip}
$a_*^{\rm{max}}/\rm{Myr}$  & $(2, 1500)$ & Uniform & Constant \\ %
 \noalign{\smallskip}
$a_*/\rm{Myr}$  & $(2, 1000)$ & Uniform & exponential \\ %
 \noalign{\smallskip}
   &  &  & delayed\\ %
    \noalign{\smallskip}
   &  &  & lognormal\\ %
 \noalign{\smallskip}
  $\tau_a/\rm{Myr}$  & $(1, 1000)$ & Uniform & exponential \\ %
 \noalign{\smallskip}
    &  &  & delayed \\ %
     \noalign{\smallskip}
    &  &  & lognormal \\ %
 \noalign{\smallskip}
\hline
\end{tabular}
\tablefoot{The actual upper limit on the prior for $a_*$, $a_*^{\rm{min}}$, $a_*^{\rm{max}}$, and $\tau$ is set to be the age of the Universe at the redshift of the galaxy.}
\end{table}

\subsection{Dust parametrization}

\texttt{BAGPIPES} includes two flexible dust attenuation models: the modified \cite{2000ApJ...539..718C} model and the \cite{2018ApJ...859...11S} model. However, the flexibility of the two models is somewhat limited, as they are not able to recover well all the conventional dust attenuation templates along with the global properties of the simulated galaxies. 
In this work, we adopt the parameterization of dust attenuation proposed by \cite{2008ApJ...685.1046L}, which is able both to recover standard dust curves (Calzetti, MW, SMC and LMC) and to consider potential non-conventional ones. The analytical expression for the dust attenuation law, normalized to the attenuation in the rest-frame optical range at $0.55 \ \mu m$ ($A_V$) is:

\begin{equation}
\begin{split}   
A_{\lambda}/A_V &= \frac{c_1}{(\lambda/0.08)^{c_2}+(0.08/\lambda)^{c_2} +c_3} \\ 
& +  \frac{233[1-c1/(6.88^{c_2}+0.145^{c_2}+c_3)-c_4/4.60]}{(\lambda/0.046)^2+(0.046/\lambda)^2+90} \\ 
& + \frac{c_4}{(\lambda/0.2175)^2+(0.2175/\lambda)^2-1.95},
\end{split}
\end{equation}%
where $c_1$, $c_2$, $c_3$ and $c_4$ are dimensionless parameters and $\lambda$ is the wavelength in $\mu m$. The three terms of the Drude model describe the far ultraviolet (FUV) attenuation rise, attenuation in the optical and near infrared (NIR) range, and the $2175 \AA$ bump (which is pronounced in the MW and, to a lesser extent, in the LMC curve), respectively. The $c_1-c_4$ parameters reproducing the Calzetti, SMC, LMC, MW templates are reported in Table \ref{table_pars}. 

\begin{table}[h]
\centering
 \caption{Drude model fit parameters for the Calzetti, SMC, MW, and the LMC templates from \cite{2008ApJ...685.1046L}.
 \label{table_pars}
 }
\begin{tabular}{lcccccccc}
 \hline \hline
  \noalign{\smallskip}
  curve & Calzetti & SMC & MW & LMC  \\
   \noalign{\smallskip}
 \hline
 \noalign{\smallskip}
$c_1$ & 44.9 & 38.7 & 14.4 & 4.47 \\
 \noalign{\smallskip}
$c_2$ & 7.56  &  3.83 & 6.52 &  2.39\\ %
 \noalign{\smallskip}
 $c_3$ & 61.2 &  6.34 & 2.04 &  -0.988 \\ %
 \noalign{\smallskip}
$c_4$ & 0& 0   & 0.0519 & 0.0221\\ %
 \noalign{\smallskip}
\hline
\end{tabular}
\end{table}

The disadvantage of this method is that dust attenuation is described by four additional parameters with respect to the conventional fitting procedure. Thus, it can be used only with spectroscopic observations (as in our case) and/or when a sufficiently large number of photometric data points are available (e.g. with the {\it JWST} Public Release IMaging for Extragalactic Research - PRIMER large program).
The advantage is that it can model any potential variation of the dust attenuation law from the ones derived from local star-forming galaxies. In fact, contrarily to the conventional fitting procedure, our method does not require to assume {\it a priori} attenuation curve. This aspect is particularly important in the case of high-$z$ galaxies since, as discussed in the Introduction, there are no physical reasons supporting the idea that dust attenuation curves at early epochs resemble any of the ones found in the local Universe. 

We implement the Drude parameterization in the \texttt{BAGPIPES} SED fitting software package. The four parameters ($c_1$, $c_2$, $c_3$ and $c_4$) of the Drude parameterization are constrained simultaneously with parameters of the SFH model and the global physical parameters of the source. The list of all the free parameters, along with their priors and ranges, is given in Table \ref{params}.
The prior probability densities and allowed ranges for the model parameters are selected based on the suggestions from the literature (\citealp{2018MNRAS.480.4379C, 2019ApJ...873...44C, 2019MNRAS.490..417C, 2022MNRAS.tmpL.128C};  \citealp{2019ApJ...876....3L}) and the tests performed in retrieving the parameters of simulated galaxies (Appendix \ref{fit_syn}).
We allow the parameters of the dust parametrization model to vary in a wide range that covers all the values of the Drude model fit to the Calzetti, SMC, LMC and MW empirical curves (Table \ref{table_pars}; see also \citealp{2008ApJ...685.1046L}), and any potential unconventional curve (Fig. \ref{curves}, gray region). 
The prior limits for the $c_4$ parameter, which characterizes the "MW bump" feature, are set so that $c_4 \gtrsim 0$, since negative values have no physical meaning.

\begin{table}[h]
\centering
 \caption{Free parameters and their priors that are used in the SED fitting procedure.
 \label{params}
 }
\begin{tabular}{lcccccccc}
 \hline \hline
  \noalign{\smallskip}
  Parameter & Limits & Prior \\
   \noalign{\smallskip}
 \hline
 \noalign{\smallskip}
 Global & & \\
  \hline
 \noalign{\smallskip}
$z$ & $(z_{\rm{sp}}-0.5, z_{\rm{sp}}+0.5)$ & Uniform \\
 $\sigma_{\rm{v}}/{\rm{km \ s^{-1}}}$ & $(1, 1000)$ & Logarithmic \\
 \noalign{\smallskip}
  \hline
 \noalign{\smallskip}
 SFH & & \\
  \hline
 \noalign{\smallskip}
$\log{M_*^{\rm{form}}/M_{\odot}}$ & $(7, 12)$ & Uniform \\
 \noalign{\smallskip}
$Z/Z_{\odot}$  & $(0.001, 1)$ & Logarithmic \\ %
 \noalign{\smallskip}
 $\Delta \log(\rm{SFR})_i$ & $(-10, 10)$ & Student's-t \\
 \noalign{\smallskip}
   \hline
 \noalign{\smallskip}
 nebular emission & & \\
  \hline
 \noalign{\smallskip}
$\log{U}$ & $(-4, 0)$ & Uniform \\ %
 \noalign{\smallskip}
 \hline
 \noalign{\smallskip}
 dust & & \\
  \hline
 \noalign{\smallskip}
$A_V/{\rm{mag}}$ & $(0, 2)$ & Uniform \\ %
 \noalign{\smallskip}
 $c_1$ & $(0, 50)$ & Uniform  \\ %
 \noalign{\smallskip}
  $c_2$ & $(0, 10)$ & Uniform \\ %
 \noalign{\smallskip}
  $c_3$ & $(-1, 75)$ & Uniform \\ %
 \noalign{\smallskip}
  $c_4$ & $(-0.005, 0.06)$ & Uniform \\ %
 \noalign{\smallskip}
\hline
\end{tabular}
\end{table}


\section{Results}\label{Results}

In this section, we fit the NIRSpec {\it JWST} spectroscopic observations described in Sect. \ref{Data} with the SED fitting method outlined in Sect. \ref{method}. In Sect. \ref{curve}, we infer the physical properties and the dust attenuation parameters of our galaxies. In Sect. \ref{comparison}, we discuss how much the derived physical properties differ from their fiducial values if we {\it a priori} assume conventional dust attenuation templates. Finally, in Sect. \ref{SFH}, we test whether our results depend on the adopted SFH model.

\subsection{Dust attenuation curves in high-$z$ star-forming galaxies} \label{curve}

We use \texttt{BAGPIPES} to load the galaxy spectra, masking the low-wavelength spectral region  below the Ly$\alpha$ ($\lambda \lesssim \lambda_{\rm{Ly\alpha}}(1+z_{\rm{sp}})$, where $z_{\rm{sp}}$ is the spectroscopic redshift; \citealp{2023arXiv230107072T}), where the flux can be attenuated by the intervening neutral IGM. Next, we provide the instructions on the model that we use to fit the data, i.e. the parameters of the model and their priors as summarized in Table \ref{params}. Using MultiNest we sample the posterior distribution in parameter space and obtain the best-fit parameters of the model with their $1\sigma$ uncertainties.

\begin{figure*}
\centering
\includegraphics[width=0.56\hsize]{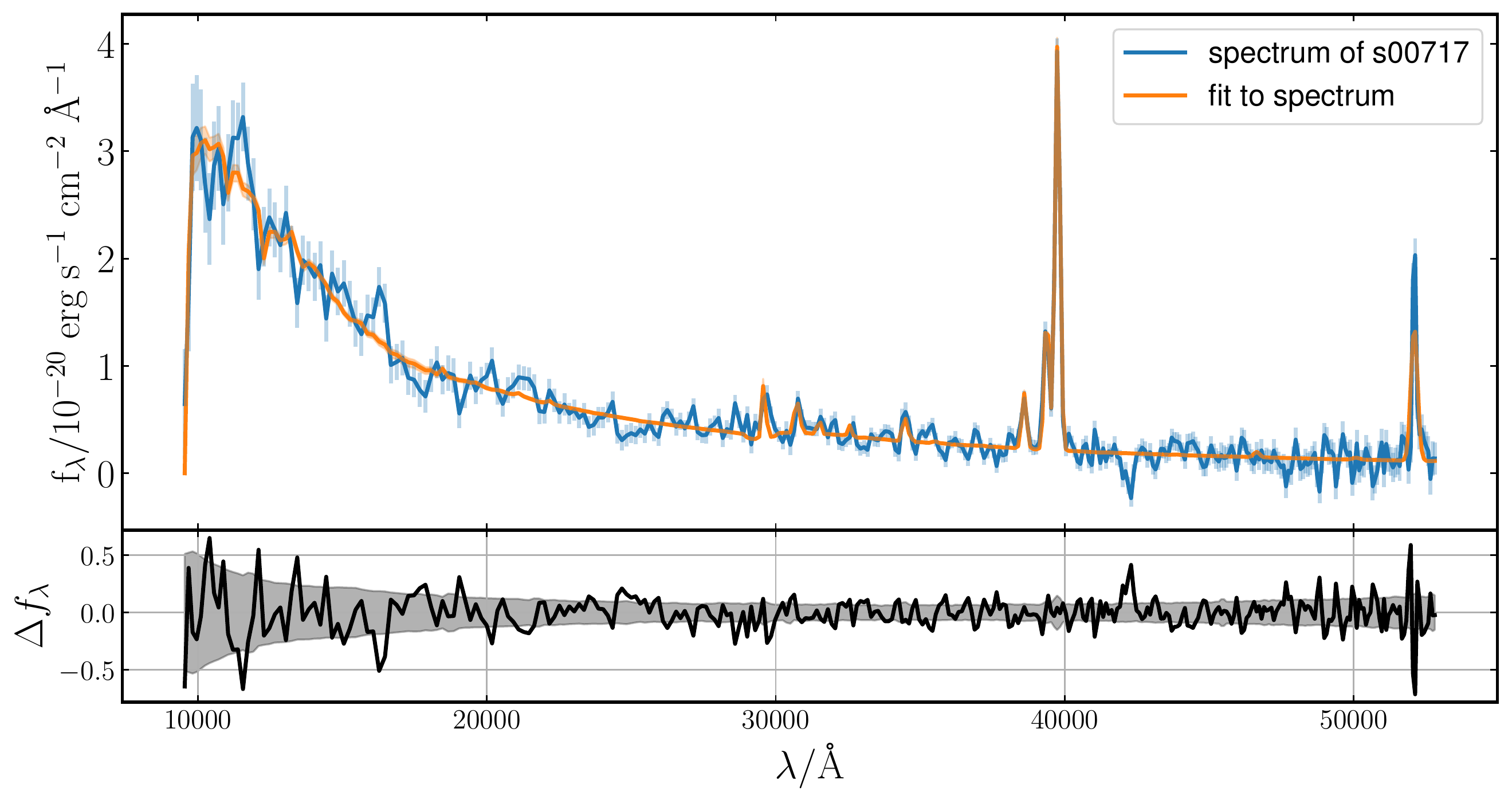}
\includegraphics[width=0.38\hsize]{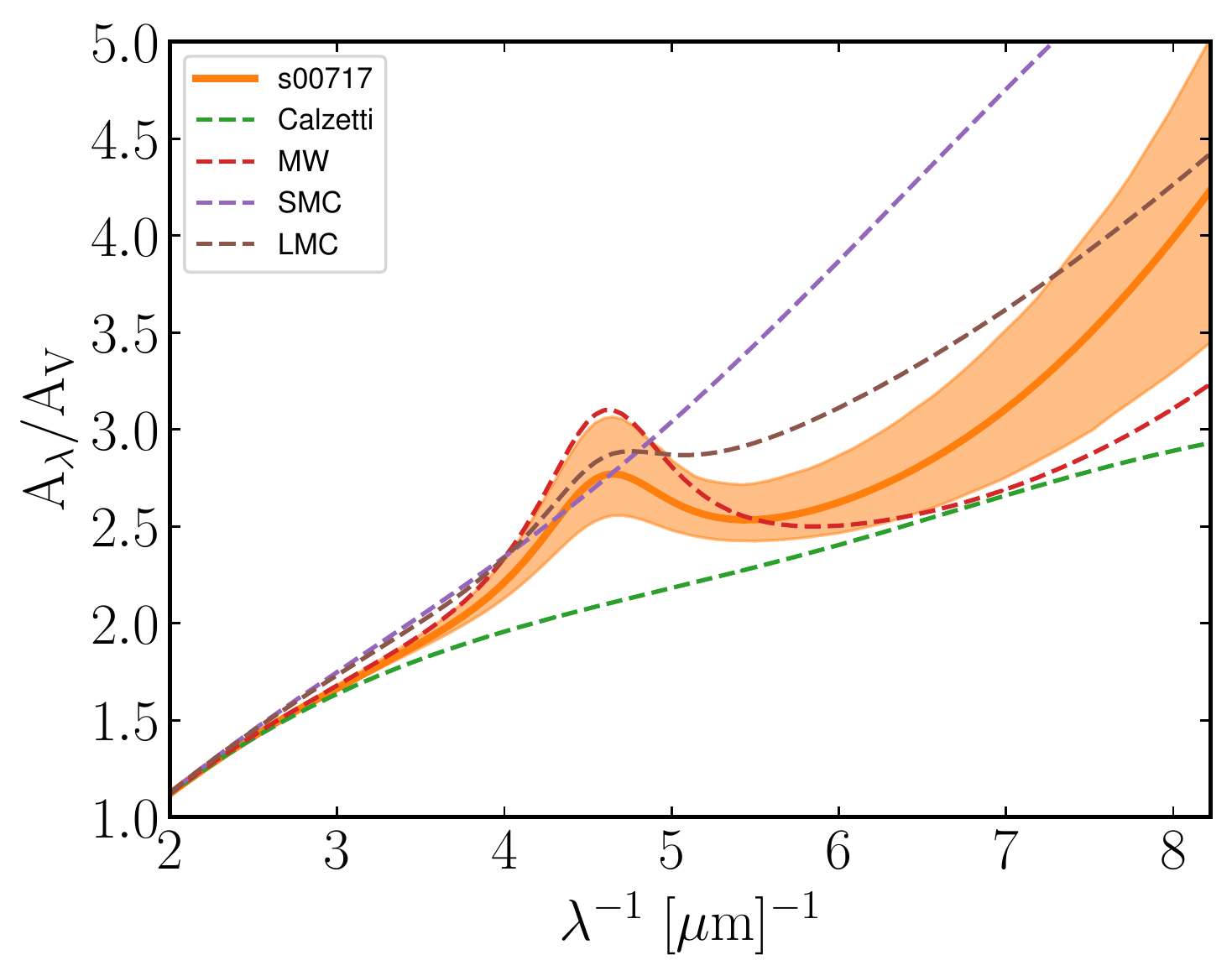}
\includegraphics[width=0.56\hsize]{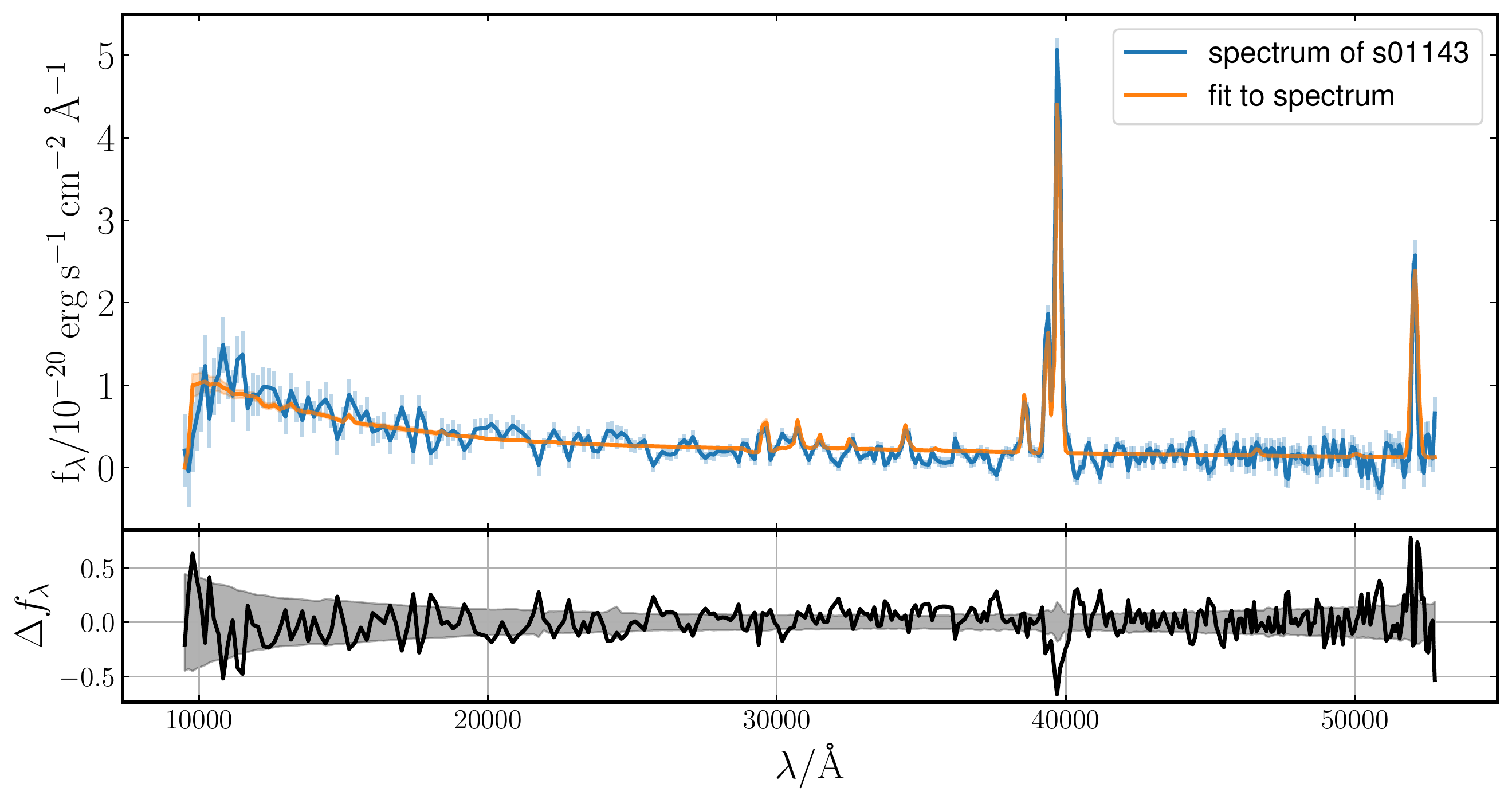}
\includegraphics[width=0.38\hsize]{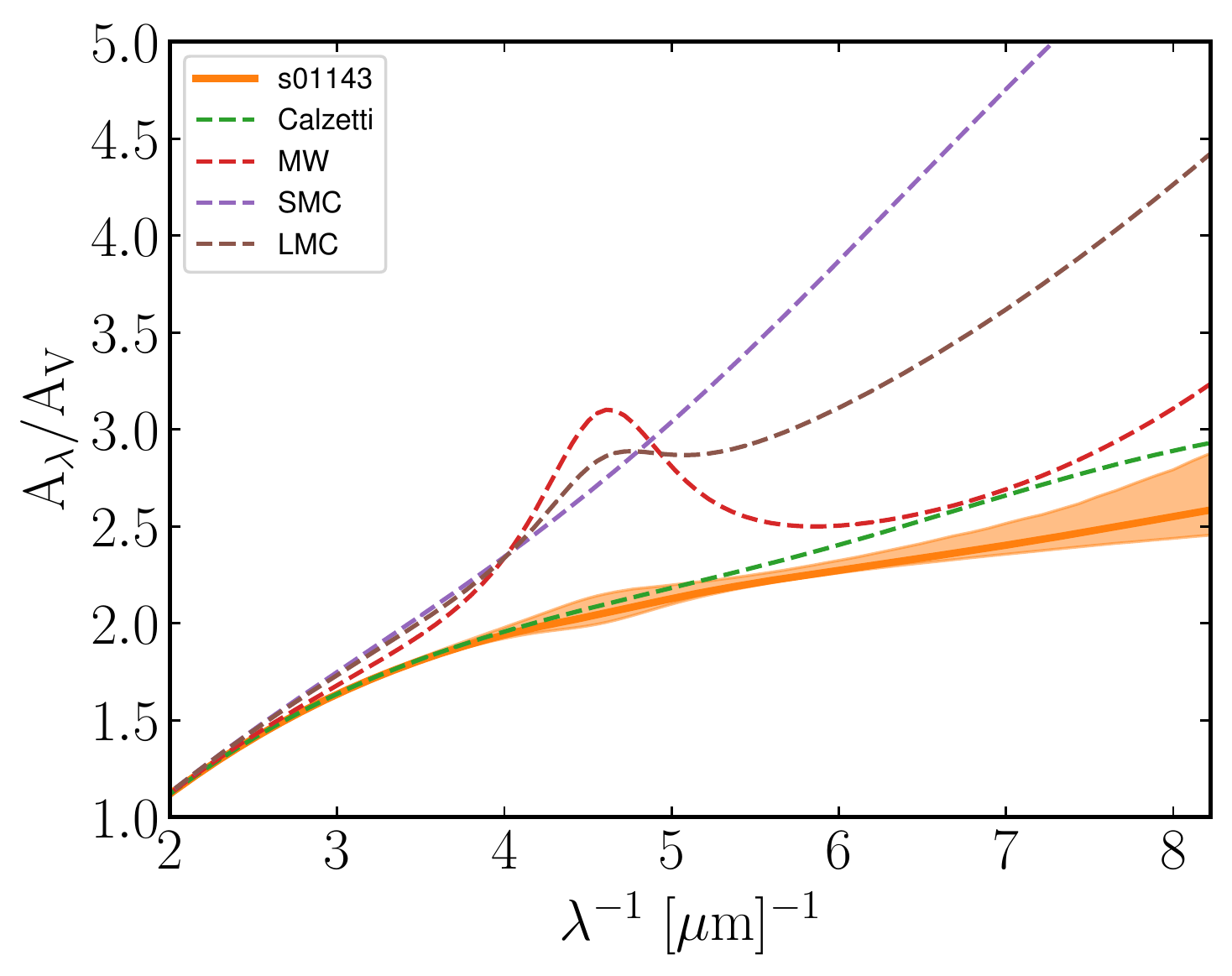}
\includegraphics[width=0.56\hsize]{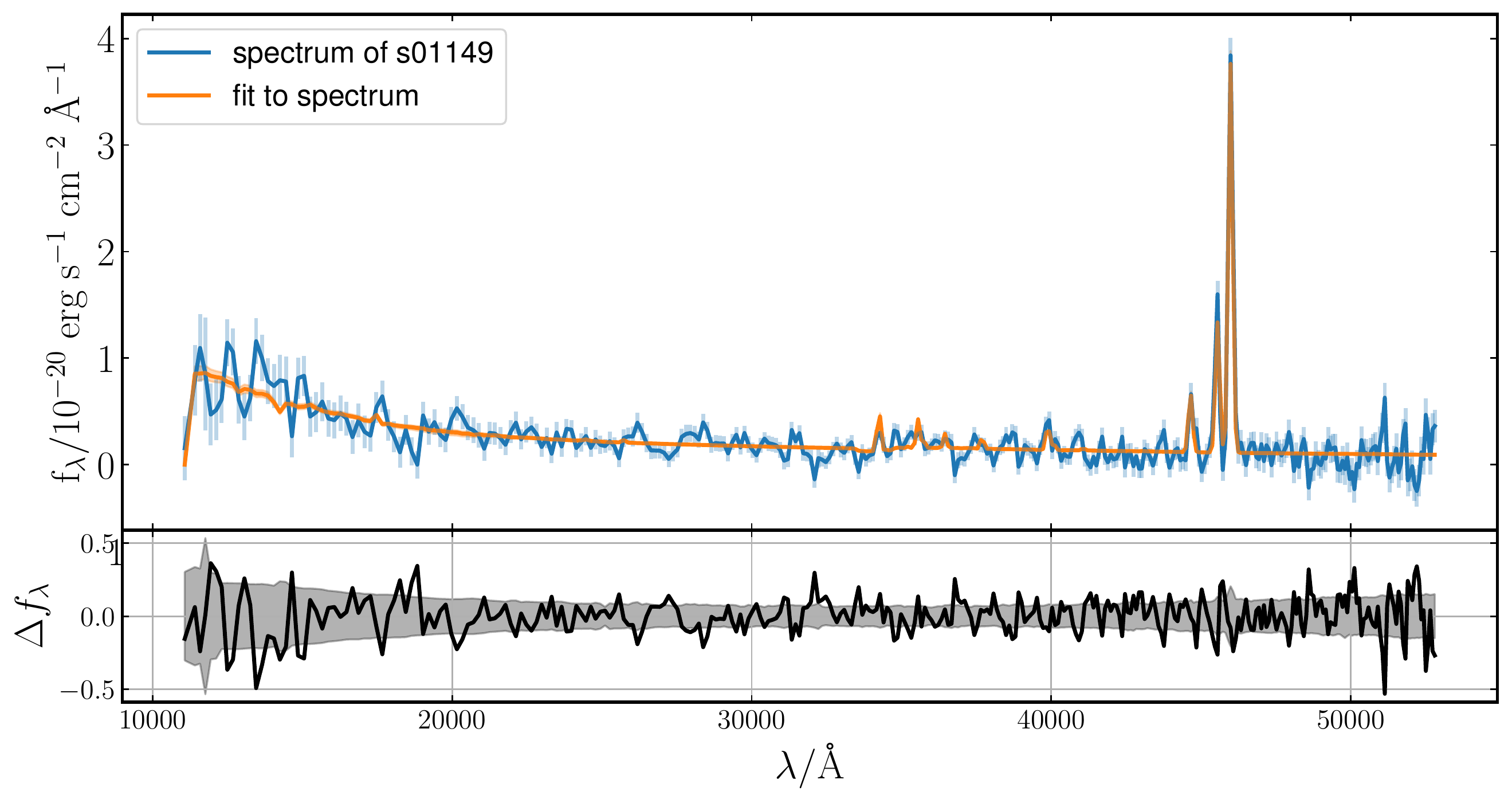}
\includegraphics[width=0.38\hsize]{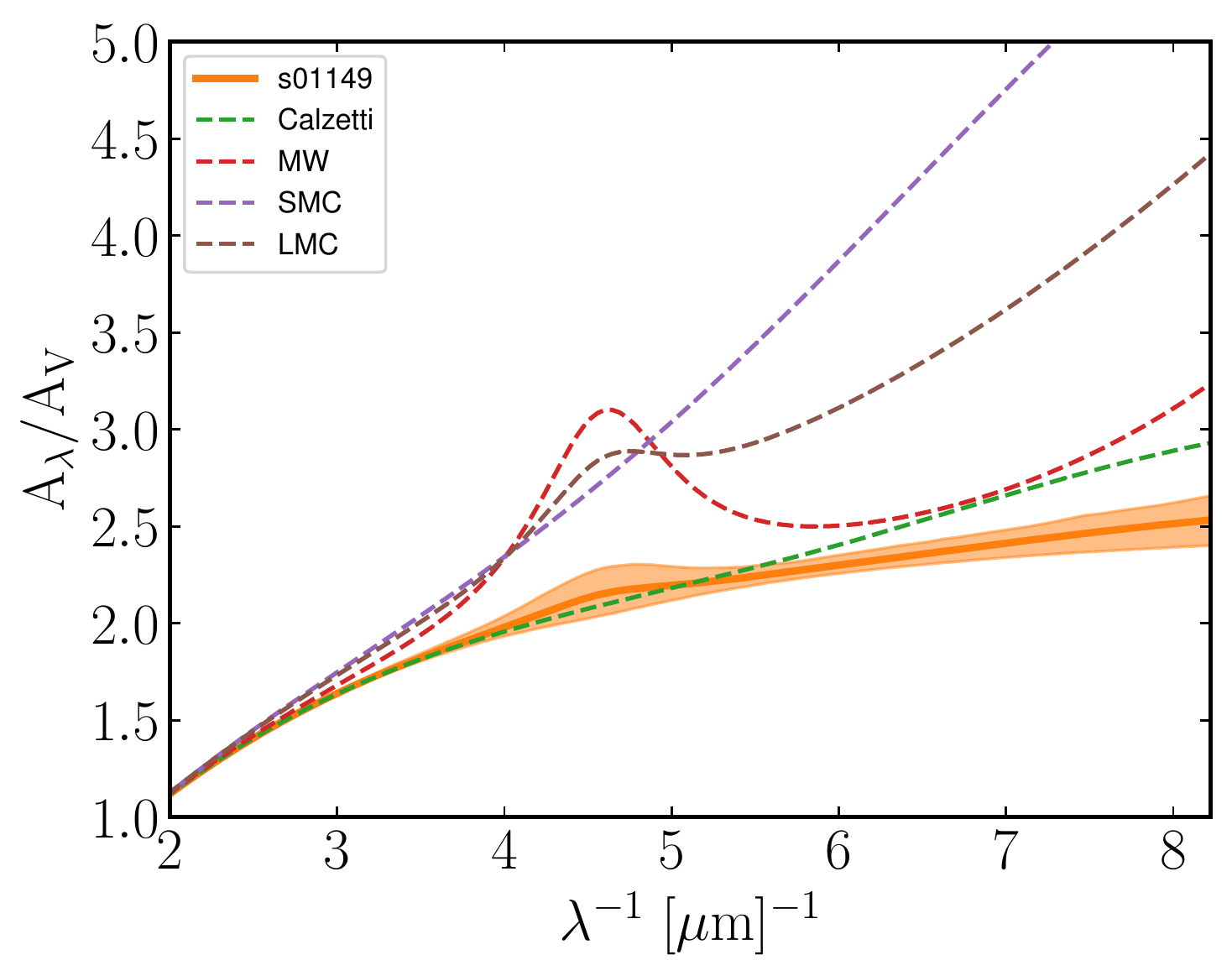}
 \caption{Top left: NIRSpec {\it JWST} spectrum of s00717 (top), s01143 (middle) and s01149 (bottom) is shown in blue, with flux uncertainties illustrated in pale blue. Orange and pale orange color indicate the best-fit posterior spectrum with $1\sigma$ uncertainties, respectively. Bottom left: The residuals of the best-fit on the observed spectra $\Delta f_{\lambda}$ with 1$\sigma$ uncertainties. Right: The best-fit dust attenuation model with $1\sigma$ uncertainties obtained using the SED fitting method with the non-parametric SFH model, for s00717, s01143, and s01149 sources.\label{spec}
 }
\end{figure*}

Fig. \ref{spec} depicts the {\it JWST} NIRSpec spectra of the three galaxies (s00717, s01143, and s01149) analyzed in this work, along with the best-fit posterior spectra (left panels, blue and orange spectra, respectively). The shaded regions represent their respective $1\sigma$ uncertainties. Although the best fit spectrum fits narrow nebular lines at longer wavelengths ($\lambda_{\rm{obs}} \gtrsim 30 000 \ \AA$); e.g., the [\ion{O}{III}] $\lambda\lambda 4959, 5007$ doublet line, H$\beta$, H$\gamma$, and [\ion{O}{II}] $\lambda 3727$), the fit to the broader potentially detected, rest-frame UV lines (such as \ion{C}{III]} $\lambda\lambda 1907, 1909$, \ion{C}{IV} $\lambda\lambda 1548, 1550$, \ion{N}{III]} $\lambda 1747$, \ion{N}{IV} $\lambda 1486$, and \ion{He}{II} $\lambda 1640$) is somewhat poorer. The broadening of the rest-frame UV lines compared to the rest-frame optical lines can be attributed to the varying spectral resolution of the observed NIRSpec spectra, which increases with the wavelength and ranges from $R \sim 50-400$. Conversely, BAGPIPES (version 1.0.0) provides output best-fit spectra at a constant resolution (default value of $R = 1000$), which is well-suited for fitting the narrow optical lines. In order to test if this affects our results, we perform fit on the observed spectra that are smoothed (i.e., convolved with a Gaussian kernel) to the lower resolution of $R \sim 100$, and we find that the inferred galaxy and dust attenuation parameters remain consistent (within $\sim 1-2\sigma$).

The dust attenuation curves derived for each source, along with the 1-$\sigma$ uncertainties are illustrated on right panels of Fig. \ref{spec} (orange solid lines with shaded regions). To estimate these uncertainties, we first boot-strap 1000 attenuation laws from a random sampling of the $c_1-c_4$ from the 1D posterior distributions. Then, we plot the median attenuation curve with $1\sigma$ dispersion of the distribution.

\begin{figure*}
\centering
\includegraphics[width=\hsize]{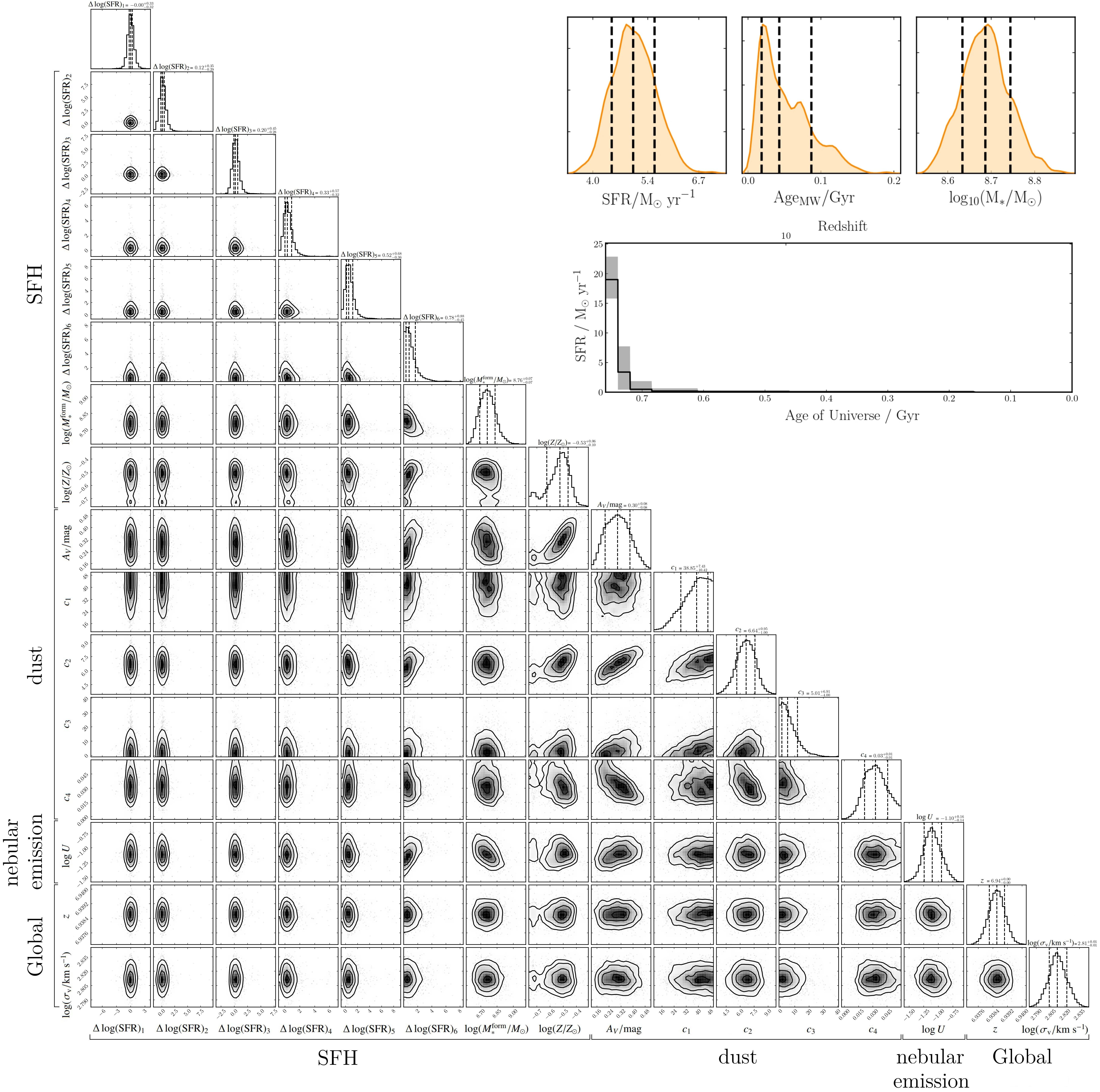}
 \caption{Corner plot illustrating the the 1D and 2D projections of the posterior distribution of the parameters derived from the SED fitting on spectrum of s00717. 1D posterior of the additional properties derived from the SFH (the SFR, the living stellar mass, and the mass-weighted age of the galaxy) and SFH are shown on the right.
 \label{corner}
 }
\end{figure*}

 We report the corner plots with the 1D and 2D projections of the posterior distribution of all the parameters fitted by our model, for s00717, s01143, and s01149 (Fig. \ref{corner}, Fig. \ref{corner2}, and Fig. \ref{corner3}, respectively)\footnote{In a few cases, the posterior distribution of certain dust attenuation parameters is close to the assumed prior limits, making them poorly determined. Nonetheless, our capacity to constrain the dust attenuation curve remains unaffected, as shown by small variations induced by the parameter uncertainty (right panels of Fig. \ref{spec}).}. Next, on the right top inset of  Fig. \ref{corner} we illustrate the 1D projection of the posterior of the global fundamental properties of the s00717 source: the living stellar mass $\log{M_*}$, the SFR, and the mass-weighted age of the galaxy ${\langle a \rangle}_*^{\rm{m}}$ that we will focus on throughout the paper. Finally, on the right bottom inset of Fig. \ref{corner}, we illustrate the best-fit SFH of s00717 (that is, the SFR as a function of the age of the Universe), modeled by the fiducial non-parametric SFH with a continuity prior. The best fit parameters of the three sources, along with the derived fundamental properties ($\log{M_*}$, SFR, and ${\langle a \rangle}_*^{\rm{m}}$) are summarized in Table \ref{params_gals}.

\begin{table}[h]
\centering
 \caption{Constrained properties of the three sources using  the SED fitting procedure.
 \label{params_gals}
 }
\begin{tabular}{lcccccccc}
 \hline \hline
  \noalign{\smallskip}
   & s00717 & s01143 & s01149 \\
   \noalign{\smallskip}
 \hline
 \noalign{\smallskip}
 Global & & \\
  \hline
 \noalign{\smallskip}
$z$ & 6.939 & 6.935 & 8.187 \\
\noalign{\smallskip}
 $\sigma_{\rm{v}}/{\rm{km \ s^{-1}}}$ & $648_{-12}^{+15}$ & $615_{-12}^{+13}$ & $474_{-15}^{+15}$ \\
 \noalign{\smallskip}
 $\log{M_*/M_{\odot}}$ & $8.69_{-0.05}^{+0.06}$ & $8.89_{-0.03}^{+0.02}$ & $8.87_{-0.05}^{+0.06}$    \\
 \noalign{\smallskip}
 $\rm{SFR}/M_{\odot} \ \rm{yr^{-1}}$   &  $ 5.1_{-0.5}^{+0.5}$ &  $8.6_{-0.6}^{+0.5}$ & $8.1_{-0.6}^{+1.0}$   \\ %
 \noalign{\smallskip}
  ${\langle a \rangle}_*^{\rm{m}}/\rm{Myr}$  & $46_{-28}^{+44}$&  $13_{-3}^{+14}$ & $13_{-3}^{+20}$  \\ %
\noalign{\smallskip}
 \noalign{\smallskip}
  \hline
 \noalign{\smallskip}
 SFH & & \\
  \hline
 \noalign{\smallskip}
$\log{M_*^{\rm{form}}/M_{\odot}}$ & $8.76_{-0.07}^{+0.07}$ & $8.93_{-0.03}^{+0.03}$ & $8.92_{-0.05}^{+0.06}$\\
 \noalign{\smallskip}
$Z/Z_{\odot}$  & $0.299_{-0.062}^{+0.045}$ & $0.203_{-0.003}^{+0.005}$ & $0.202_{-0.012}^{+0.016}$ \\ %
 \noalign{\smallskip}
 $\Delta \log(\rm{SFR})_1$ & $0.00_{-0.32}^{+0.33}$  & $-0.01_{-0.35}^{+0.36}$ & $0.02_{-0.34}^{+0.35}$\\
 \noalign{\smallskip}
  $\Delta \log(\rm{SFR})_2$ & $0.12_{-0.29}^{+0.35}$ &  $0.03_{-0.32}^{+0.36}$ & $0.04_{-0.32}^{+0.38}$\\
 \noalign{\smallskip}
  $\Delta \log(\rm{SFR})_3$ & $0.20_{-0.28}^{+0.45}$ &  $0.04_{-0.32}^{+0.40}$ & $0.08_{-0.29}^{+0.41}$\\
 \noalign{\smallskip}
  $\Delta \log(\rm{SFR})_4$ & $0.33_{-0.32}^{+0.57}$ &  $0.08_{-0.32}^{+0.44}$ & $0.11_{-0.30}^{+0.44}$\\
 \noalign{\smallskip}
  $\Delta \log(\rm{SFR})_5$ & $0.52_{-0.36}^{+0.68}$ &  $0.09_{-0.31}^{+0.41}$ & $0.17_{-0.32}^{+0.47}$\\
 \noalign{\smallskip}
  $\Delta \log(\rm{SFR})_6$ & $0.78_{-0.45}^{+0.88}$ & $3.98_{-1.12}^{+2.21}$ & $2.62_{-0.92}^{+1.99}$ \\
 \noalign{\smallskip}
   \hline
 \noalign{\smallskip}
 nebular emission & & \\
  \hline
 \noalign{\smallskip}
$\log{U}$ &  $-1.10_{-0.14}^{+0.16}$ & $-1.58_{-0.09}^{+0.06}$ & $-1.43_{-0.09}^{+0.12}$ \\ %
 \noalign{\smallskip}
 \hline
 \noalign{\smallskip}
 dust & & \\
  \hline
 \noalign{\smallskip}
$A_V/{\rm{mag}}$ & $0.30_{-0.08}^{+0.08}$  & $1.22_{-0.05}^{+0.05}$  & $1.07_{-0.08}^{+0.09}$ \\ %
 \noalign{\smallskip}
 $c_1$ & $38.85_{-10.42}^{+7.43}$ & $14.62_{-10.05}^{+15.20}$ & $9.15_{-6.62}^{+11.90}$ \\ %
 \noalign{\smallskip}
  $c_2$ & $6.64_{-1.00}^{+0.95}$ & $8.96_{-1.94}^{+0.77}$ & $7.51_{-5.45}^{+1.86}$ \\ %
 \noalign{\smallskip}
  $c_3$ & $5.01_{-4.00}^{+6.91}$ & $36.14_{-25.76}^{+25.35}$ & $45.28_{-25.29}^{+18.77}$ \\ %
 \noalign{\smallskip}
  $c_4$ &$0.03_{-0.02}^{+0.02}$ & $-0.001_{-0.003}^{+0.004}$ & $0.004_{-0.005}^{+0.006}$ \\ %
 \noalign{\smallskip}
\hline
\end{tabular}
\end{table}

\begin{figure}
\centering
\includegraphics[width=\hsize]{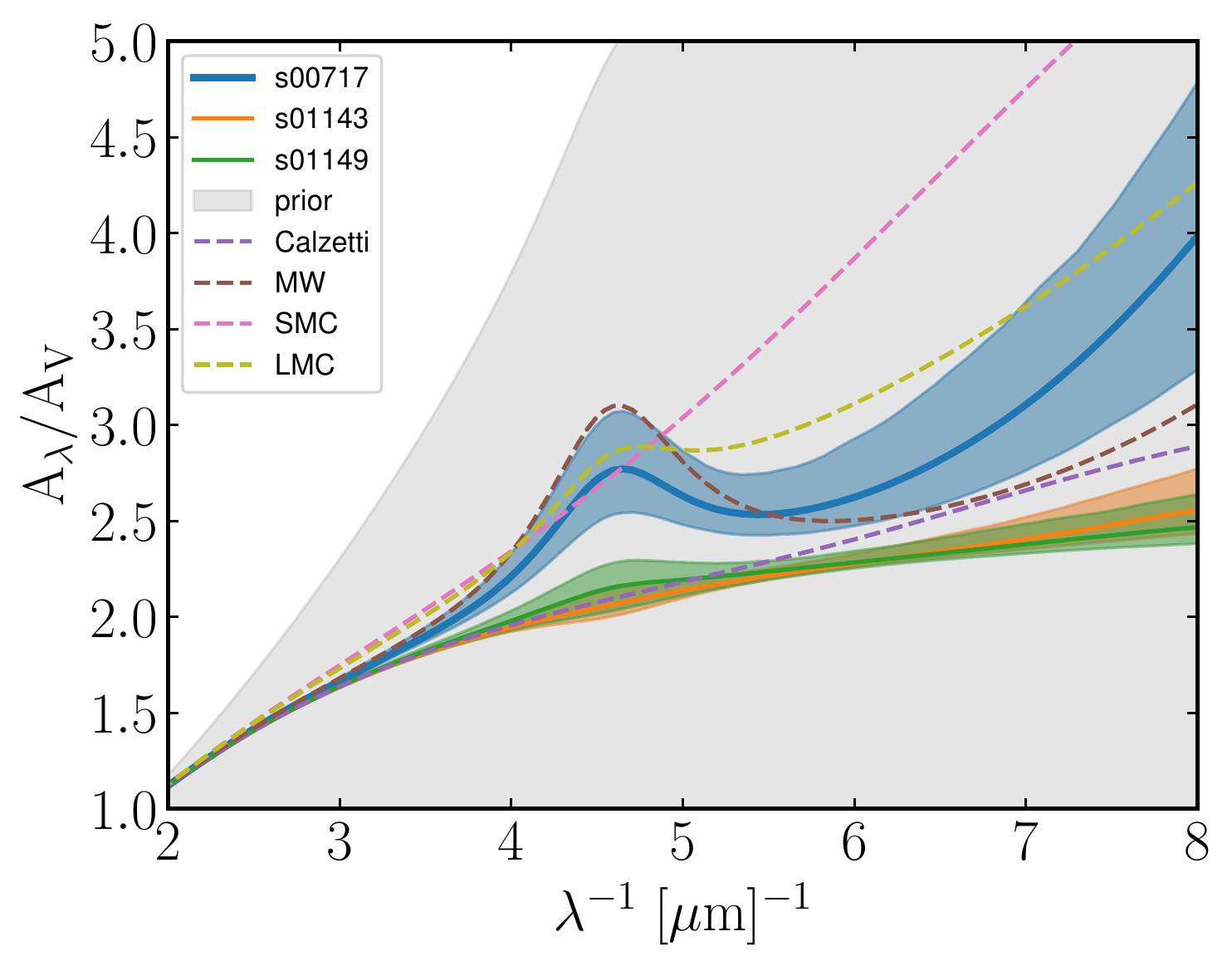}
\caption{The best-fit dust attenuation model obtained using the SED fitting method with the non-parametric SFH model, for our three sources. The median of the posterior with $1\sigma$ uncertainties are shown as solid blue, orange and green lines with associated shaded regions for s00717, s01143, and s01149, respectively. The 3$\sigma$ dispersion of the prior is estimated from a $99.7\%$ sample of 1000 random draws from the prior limits for $c_1-c_4$  parameters of the Drude model (gray shaded region). Fits to conventional dust empirical curves, frequently adopted as standard templates: the Calzetti, the MW, the SMC, and the LMC curve are shown as purple, brown, pink and yellow dashed lines, respectively.
\label{curves}
}
\end{figure}

For the three sources analyzed in this work, we report in Table \ref{params_gals} the $c_1-c_4$ parameters of the best fitting attenuation curves. By comparing these results with Table \ref{table_pars} we find that the three high-$z$ galaxies observed with JWST are characterized by dust properties different from local galaxies. Fig. \ref{curves} further highlights this result: the overall shape of the inferred attenuation curve of s00717 resembles the LMC and MW curve, but it is steeper at lower wavelengths (Fig. \ref{curves}, blue curve).
The strength of the broad UV absorption feature at $2175 \AA$ (characterized by the $c_4$ parameter of the analytical dust model), falls between that of the MW and the LMC. This distinctive feature is often linked to the Polycyclic Aromatic Hydrocarbons (PAHs, \citealp{1965ApJ...142.1683S}; see \citealp{2003ARA&A..41..241D} for a review). The identification of the $2175 \AA$ feature represents one of the first evidences of the possible presence of the small carbonaceous dust grains and/or PAHs in the spectra of a galaxy at the EoR (see also \citealp{2023arXiv230205468W}). The inferred dust attenuation curves of the remaining two sources, s01143 and s01149, are  similar to the Calzetti, but slightly flatter at short wavelengths ($\lambda \lesssim 0.2$ microns). Their dust attenuation laws show no significant evidence of the characteristic MW bump (Fig. \ref{curves}, orange and green curves). In Sect. \ref{curve_evolution}, we will delve into the inferences of these results.

\subsection{Comparison with the standard attenuation templates} \label{comparison}

In this subsection, we investigate whether the derived physical properties of galaxies change if we {\it a priori} adopt one of the standard attenuation models as a template instead of the fiducial attenuation curve that is constrained using the Drude-type parameterization. 

We thus carry out multiple \texttt{BAGPIPES} runs in order to perform the SED fitting of our spectroscopic data, as described in Sect. \ref{method}, but instead of using the Drude approach, we assume one of the conventional dust attenuation templates. Dust empirical models that have been frequently {\it a priori} adopted in the literature to infer the properties of high-$z$ galaxies are the Calzetti, the SMC, and the MW curve (\citealp{2022arXiv220711217A}; \citealp{2022arXiv220711135L}; \citealp{2022MNRAS.516..975T}; \citealp{2022arXiv220714265T}). An overview of the derived physical properties of the three galaxies, obtained with the fiducial attenuation curve, and by assuming one of the standard attenuation templates, is shown in Table \ref{table_curves_all}.

\begin{table*}[h]
\centering
\caption{A summary of the constrained physical properties of the three galaxies with different attenuation curves used in the SED fitting procedure assuming the fiducial non-parametric SFH model with a continuity prior.
\label{table_curves_all}
}
\begin{tabular}{lcccccccc}
 \hline \hline
  \noalign{\smallskip}
  Parameter & Fiducial & Calzetti & SMC &  MW  & $|\Delta|/\%$ \\
   \noalign{\smallskip}
 \hline
 \noalign{\smallskip}
\multicolumn{6}{c} {s00717}   \\
 \hline
 \noalign{\smallskip}
 $\log{M_*/M_{\odot}}$ & $8.69_{-0.05}^{+0.06}$ & $8.76_{-0.06}^{+0.06}$ & $8.72_{-0.06}^{+0.08}$ & $8.72_{-0.05}^{+0.06}$ & $7-17$ \\
 \noalign{\smallskip}
 $\rm{SFR}/M_{\odot} \ \rm{yr^{-1}}$   &  $ 5.1_{-0.5}^{+0.5}$ &  $5.8_{-0.6}^{+0.6}$ & $4.6_{-0.4}^{+0.5}$  & $5.4_{-0.4}^{+0.5}$ & $6-14$\\ 
 \noalign{\smallskip}
  ${\langle a \rangle}_*^{\rm{m}}/\rm{Myr}$  & $46_{-28}^{+44}$&  $53_{-30}^{+56}$ & $102_{-54}^{+67}$ & $46_{-26}^{+54}$ & $<122$ \\ 
 \noalign{\smallskip}
$Z/Z_{\odot}$  & $ 0.299_{-0.062}^{+0.045}$ &  $ 0.302_{-0.039}^{+0.036}$ & $ 0.195_{-0.009}^{+0.027}$  & $ 0.335_{-0.038}^{+0.036}$ & $1-35$\\ 
 \noalign{\smallskip}
$\log{U}$ & $-1.10_{-0.14}^{+0.16}$& $-1.13_{-0.14}^{+0.16}$& $-1.25_{-0.15}^{+0.14}$& $-1.18_{-0.12}^{+0.11}$ & $7-30$ \\ 
 \noalign{\smallskip}
$A_V/{\rm{mag}}$ & $0.30_{-0.08}^{+0.08}$ &   $0.39_{-0.05}^{+0.05}$ &  $0.14_{-0.02}^{+0.01}$ & $0.34_{-0.04}^{+0.04}$ & $13-53$ \\  
 \noalign{\smallskip}
\hline
\noalign{\smallskip}
\multicolumn{6}{c} {s01143}   \\
 \hline
 \noalign{\smallskip}
$\log{M_*/M_{\odot}}$ & $8.89_{-0.03}^{+0.02}$ & $8.84_{-0.02}^{+0.03}$ & $8.57_{-0.01}^{+0.03}$  & $8.75_{-0.02}^{+0.02}$ & $11-52$ \\
 \noalign{\smallskip}
 $\rm{SFR}/M_{\odot} \ \rm{yr^{-1}}$ & $ 8.6_{-0.6}^{+0.5}$ &  $7.7_{-0.4}^{+0.5}$ & $4.2_{-0.2}^{+0.2}$ & $6.2_{-0.3}^{+0.3}$ & $10-51$ \\
 \noalign{\smallskip}
  ${\langle a \rangle}_*^{\rm{m}}/\rm{Myr}$  & $13_{-3}^{+4}$&  $11_{-1}^{+3}$ & $10_{-0}^{+5}$ & $10_{-0}^{+5}$ &  $15-23$\\
 \noalign{\smallskip}
$Z/Z_{\odot}$  & $ 0.203_{-0.003}^{+0.005}$ &  $ 0.202_{-0.002}^{+0.005}$ & $ 0.201_{-0.001}^{+0.003}$ & $ 0.202_{-0.002}^{+0.005}$ & $0.5-1$\\ 
 \noalign{\smallskip}
$\log{U}$ & $-1.58_{-0.09}^{+0.06}$& $-1.59_{-0.09}^{+0.07}$& $-1.68_{-0.11}^{+0.08}$& $-1.62_{-0.09}^{+0.08}$& $2-21$ \\
 \noalign{\smallskip}
$A_V/{\rm{mag}}$ & $1.22_{-0.05}^{+0.05}$ & $1.12_{-0.04}^{+0.04}$ &  $0.59_{-0.03}^{+0.03}$ & $0.92_{-0.03}^{+0.03}$ & $8-52$ \\ 
 \noalign{\smallskip}
\hline
 \noalign{\smallskip}
\multicolumn{6}{c} {s01149}   \\
 \hline
 \noalign{\smallskip}
$\log{M_*/M_{\odot}}$ & $8.87_{-0.05}^{+0.06}$ & $8.81_{-0.04}^{+0.06}$ & $8.52_{-0.02}^{+0.03}$  &  $8.71_{-0.03}^{+0.04}$ & $13-55$ \\%
 \noalign{\smallskip}
  $\rm{SFR}/M_{\odot} \ \rm{yr^{-1}}$ & $ 8.1_{-0.6}^{+1.0}$ &  $7.1_{-0.6}^{+0.7}$ & $3.7_{-0.2}^{+0.2}$ & $5.6_{-0.3}^{+0.4}$ & $12-54$\\ 
 \noalign{\smallskip}
  ${\langle a \rangle}_*^{\rm{m}}/\rm{Myr}$  & $13_{-3}^{+20}$&  $14_{-4}^{+17}$ & $12_{-2}^{+11}$ & $13_{-3}^{+17}$ & $8$\\ %
 \noalign{\smallskip}
$Z/Z_{\odot}$  & $ 0.202_{-0.012}^{+0.016}$ &  $ 0.202_{-0.009}^{+0.014}$ & $ 0.201_{-0.003}^{+0.007}$ & $ 0.203_{-0.008}^{+0.013}$ & $0.5$\\ %
 \noalign{\smallskip}
$\log{U}$ & $-1.43_{-0.09}^{+0.12}$& $-1.45_{-0.07}^{+0.11}$& $-1.50_{-0.05}^{+0.06}$ & $-1.46_{-0.06}^{+0.11}$ & $5-15$\\ 
 \noalign{\smallskip}
 $A_V/{\rm{mag}}$ & $1.07_{-0.08}^{+0.09}$ & $0.98_{-0.05}^{+0.05}$ &  $0.45_{-0.03}^{+0.03}$ & $0.77_{-0.04}^{+0.04}$ & $8-58$ \\ %
 \noalign{\smallskip}
\hline
\end{tabular}
\end{table*}

We find that properties such as the redshift $z$, $\sigma_{\rm{vel}}$ (related to the width of the spectral lines/features due to velocity dispersion and instrumental broadening; e.g., \citealp{2018MNRAS.480.4379C}), the mass-weighted stellar age ${\langle a \rangle}_*^{\rm{m}}$, metallicity $Z$, and ionization parameter $\log{U}$ are practically independent (within $\sim 1\sigma$) on the assumed dust model, for all the three sources. The only exception is the inferred metallicity of the s00717 source when adopting the SMC curve, which is lower from its fiducial value by $\sim 4 \sigma$ (0.18 dex). The remaining properties instead may differ by $\sim 1-2 \sigma$ (in some cases even $>5 \sigma$), depending on the adopted attenuation curve. 

If a Calzetti dust attenuation curve is assumed, most of the derived properties of the three galaxies are within $1-2\sigma$ from the fiducial value. Next, if the MW curve is assumed, the properties of s00717 are within $\sim 1\sigma$ of their fiducial values, since the fiducial attenuation curve of s00717 resembles that of the MW (Fig. \ref{curves}, blue solid line). However, for the remaining two sources, properties such as the living stellar mass $\log{M_*}$, SFR and $A_V$ deviate significantly from their fiducial values. For s01143, these three parameters are lower by $7\sigma$ (0.13 dex), $8\sigma$ (0.14 dex), and $10\sigma$ (0.12 dex), respectively. For s01149, $\log{M_*}$, SFR and $A_V$ are lower by $4 \sigma$ (0.16 dex), $6 \sigma$ (0.16 dex), and $7.5 \sigma$ (0.14 dex), respectively. 

Finally, when assuming the SMC curve, most of the properties of s00717 resemble their fiducial values apart from $A_V$ which is lower by $> 10\sigma$ (0.33 dex). For the remaining two sources, $\log{M_*}$, SFR and $A_V$ deviate substantially ($> 10\sigma$) from their fiducial values, since their fiducial attenuation curve is much flatter than the SMC curve (Fig. \ref{curves}, green and orange solid lines for s01143 and s01149, respectively). For example, for the s01143 and s01149 sources, $\log{M_*}$, SFR and $A_V$ are lower by by $0.32-0.35$ dex, $0.31-0.34$ dex, and $0.32-0.38$ dex, respectively. 

\begin{figure*}
\centering
\textbf{Variation of galaxy parameters with adopted dust attenuation curve}\par\medskip
\includegraphics[width=0.33\hsize]{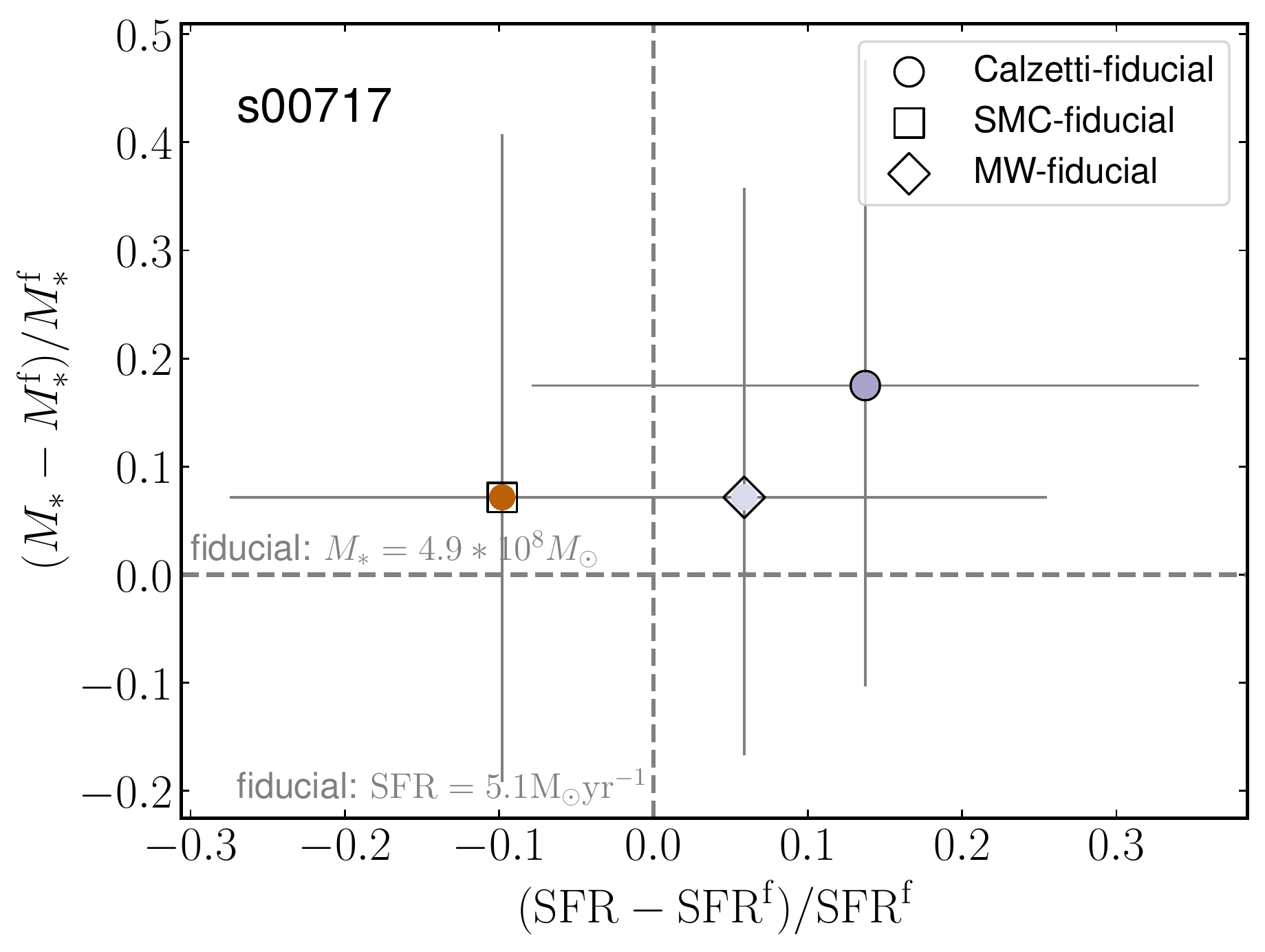}
\includegraphics[width=0.33\hsize]{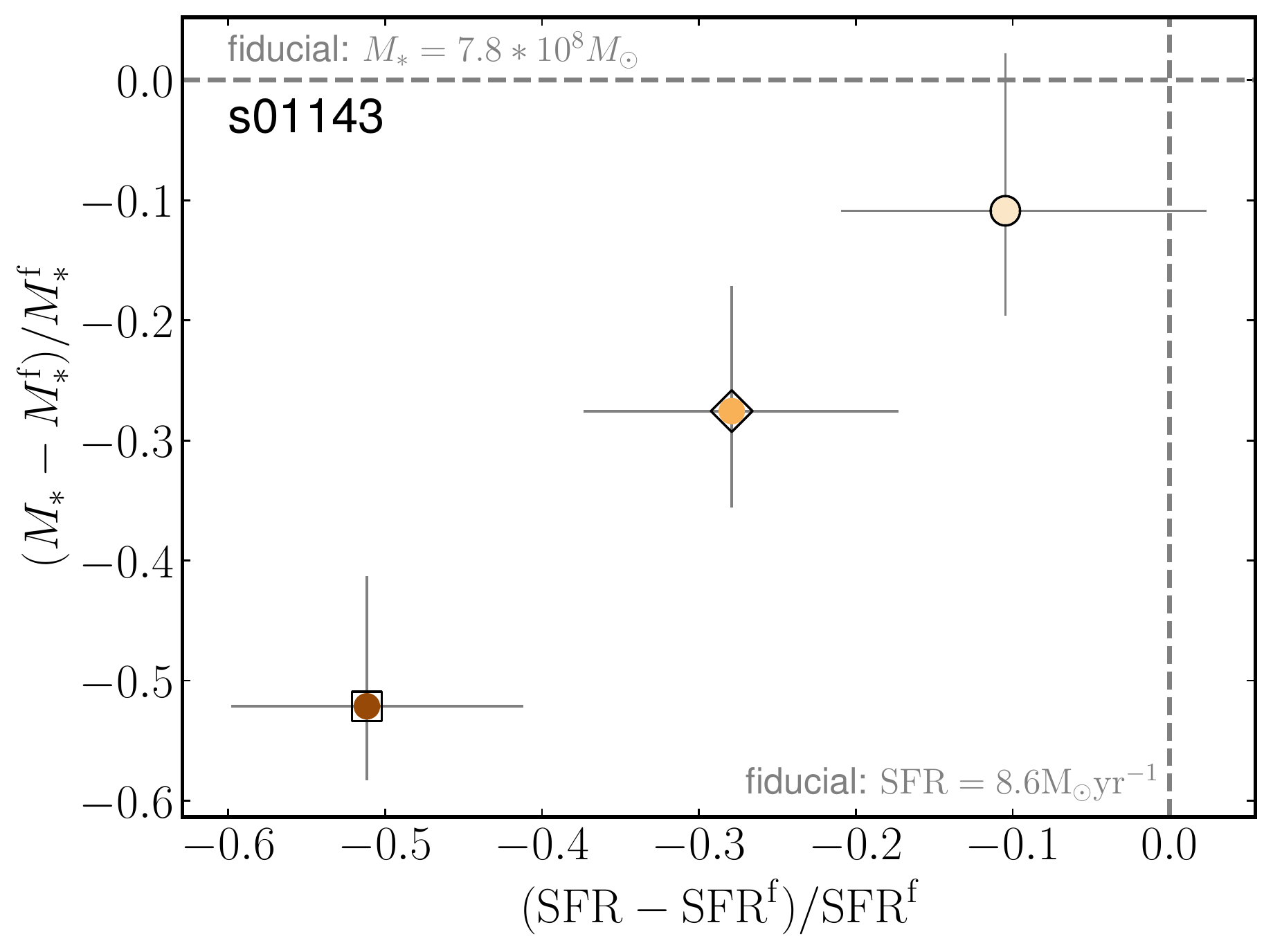}
\includegraphics[width=0.33\hsize]{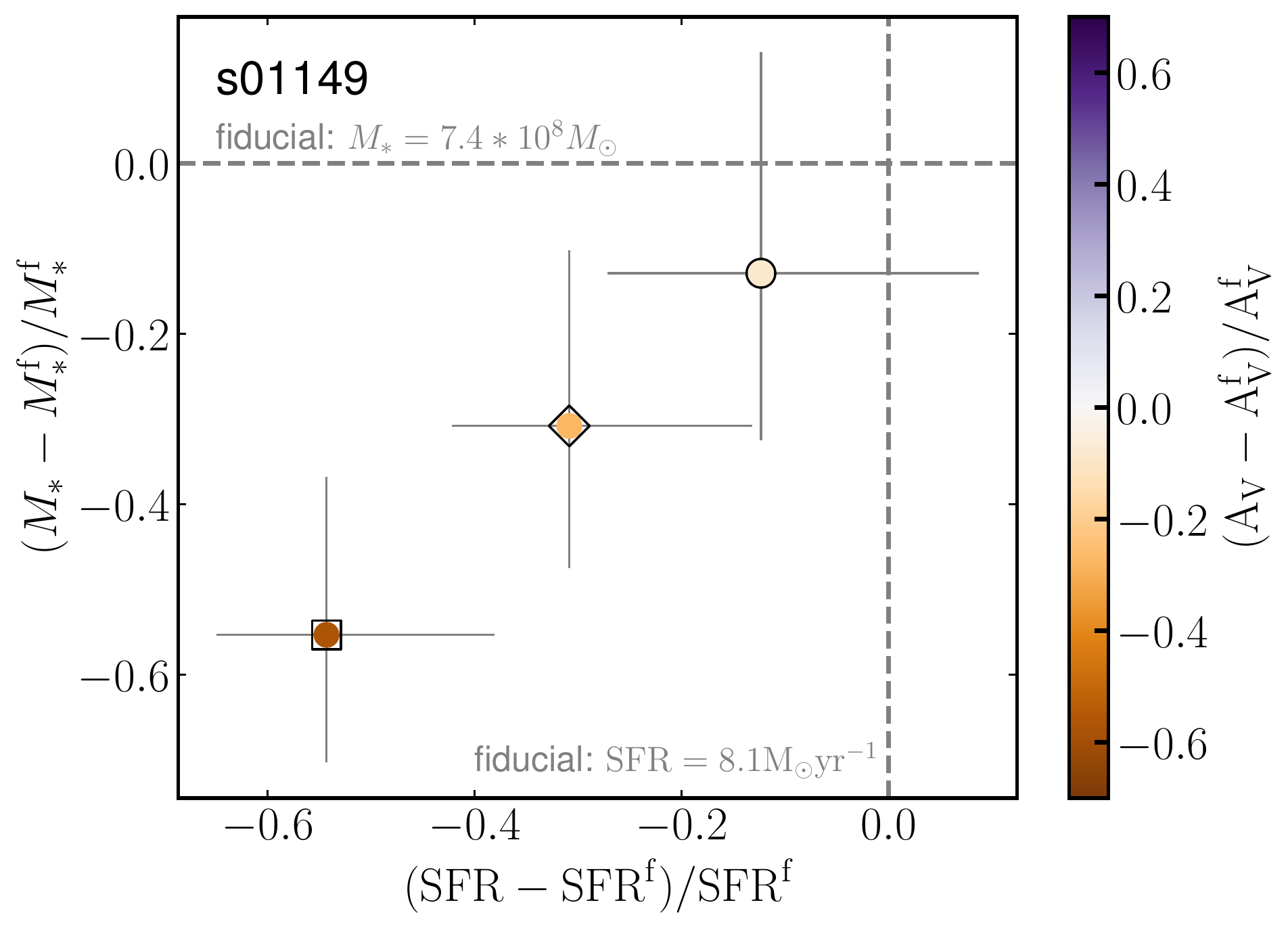}
\caption{Comparison of galaxy parameters: stellar mass ${M_*}$ and SFR, color-coded by the dust attenuation parameter $\rm{A_V}$ inferred when adopting one of the standard attenuation templates, and with the fiducial dust attenuation curve for s00717 (left), s01143 (middle), and s01149 (right). The relative variation in parameters inferred adopting the Calzetti, SMC and MW curve with the fiducial curve are represented as circle, square, and diamond, respectively.}
     \label{par_var_curve}
\end{figure*}

We summarize the aforementioned findings in Fig. \ref{par_var_curve}. Overall, the largest deviation is found for the for s01143 and s01149 for which the parameters $\log{M_*}$ and $A_V$ diverge the most from their fiducial values adopting the SMC curve. This is unsurprising given that these sources have a much flatter fiducial attenuation curve than the steep SMC extinction curve (Fig. \ref{curves}, orange and green curves, respectively). We will discuss the implications of these findings in Sect. \ref{dust_discussion}.

Lastly, in order to demonstrate that the fiducial model with the analytical dust attenuation curve provides a better fit to the observed spectra, with respect to the models with adopted standard curves, we use Bayesian model selection through the Bayes factor (\citealp{jeffreys1983theory}; \citealp{10.2307/2291091}; \citealp{10.2307/2291387}; \citealp{MacKay2003}). Bayes factor is the ratio of the marginal likelihoods or model evidences $P(D|M)$ of the two competing models $M_1$ and $M_2$, given the data $D$, and it is defined as $BF_{2-1} = P(D|M_2)/P(D|M_1)$. 

Table \ref{table_Bayes} presents the logarithm of Bayes factors obtained through the model evidences provided by the MultiNest nested sampling method (\citealp{2008MNRAS.384..449F, 2019OJAp....2E..10F}) utilized in \texttt{BAGPIPES}. Model selection based on the Bayes factor favors the fiducial model over any of the models that rely on a {\it a priori} adoption of one of the standard dust laws. In fact, the Bayes factors rank as either strong ($\log(BF) = 1-2$) or decisive ($\log(BF) > 2$), according to the proposed scale of \cite{10.2307/2291091}. 

\begin{table}[h]
\centering
 \caption{The logarithm of the Bayes factor estimates derived comparing the fiducial fitting model with the analytical dust curve and the models with adopted Calzetti, SMC, and MW templates.
 \label{table_Bayes}
 }
\begin{tabular}{lcccccccc}
 \hline \hline
  \noalign{\smallskip}
 $\log(BF_{\rm{2-1}})$ & s00717 & s01143 & s01149 \\
 \hline
 \noalign{\smallskip}
Fiducial-Calzetti & 1.72 & 1.48 & 1.92 \\
 \noalign{\smallskip}
 \hline
 \noalign{\smallskip}
Fiducial-MW  & 3.29 & 9.46 & 4.11 \\
 \noalign{\smallskip}
\hline
 \noalign{\smallskip}
Fiducial-SMC  & 1.79 & 30.74 & 15.44 \\
 \noalign{\smallskip}
\hline
\end{tabular}
\end{table}

\subsection{Comparison with different SFH models}\label{SFH}

We have so far assumed as fiducial the non-parametric SFH model with a continuity prior. In this subsection, we investigate whether the inferred galaxy physical properties and the dust attenuation law depend on the assumed SFH model.  
We run the \texttt{BAGPIPES} code and perform the SED fitting of the spectroscopic data, adopting the Drude parametrization for the dust attenuation (Sect. \ref{method}), and assuming either one of the parametric SFH models (constant, double power-law, exponential, delayed, and lognormal) or the non-parametric SFH model with a "bursty continuity" prior. We vary the free parameters of the aforementioned parametric and non-parametric SFH models, as summarized in Table \ref{params_limits} and Table \ref{params}, respectively. 

\begin{table*}[h]
\centering
\caption{A summary of constrained parameters of the source with the SED fitting procedure using different SFH models, assuming the Drude-type dust attenuation law.
\label{table_SFH}
}
\begin{tabular}{lcccccccc}
 \hline \hline
  \noalign{\smallskip}
  Parameter & Continuity &  Bursty cont. & Constant & DPL & Exponential & Delayed & Lognormal & $|\Delta|/\%$ \\
   \noalign{\smallskip}
 \hline
 \noalign{\smallskip}
\multicolumn{9}{c} {s00717}   \\
 \hline
 \noalign{\smallskip}
 $\log{M_*/M_{\odot}}$ & $8.69_{-0.05}^{+0.06}$ & $8.65_{-0.04}^{+0.04}$ & $8.43_{-0.08}^{+0.08}$ & $8.66_{-0.05}^{+0.05}$ &  $8.65_{-0.05}^{+0.04}$  & $8.69_{-0.04}^{+0.05}$ & $8.74_{-0.03}^{+0.03}$ & $<45$ \\
 \noalign{\smallskip}
 $\rm{SFR}/M_{\odot} \ \rm{yr^{-1}}$   &  $ 5.1_{-0.5}^{+0.5}$ &   $ 5.0_{-0.5}^{+0.4}$  & $2.9_{-0.5}^{+0.6}$ & $ 5.2_{-0.6}^{+0.7}$ & $ 5.1_{-0.6}^{+0.6}$ &  $5.8_{-0.7}^{+0.6}$  & $6.3_{-0.3}^{+0.4}$ & $<43$\\ %
 \noalign{\smallskip}
  ${\langle a \rangle}_*^{\rm{m}}/\rm{Myr}$  & $46_{-28}^{+44}$&  $15_{-5}^{+14}$ & $5_{-1}^{+3}$ & $ 15_{-2}^{+5}$ & $14_{-3}^{+4}$ &  $19_{-7}^{+6}$ & $32_{-6}^{+7}$ & $30-89$ \\ %
 \noalign{\smallskip}
$Z/Z_{\odot}$  & $ 0.299_{-0.062}^{+0.045}$ &  $ 0.329_{-0.050}^{+0.041}$ & $ 0.273_{-0.049}^{+0.048}$ &$ 0.275_{-0.071}^{+0.064}$ & $  0.272_{-0.075}^{+0.061}$ &$ 0.300_{-0.088}^{+0.136}$ & $0.282_{-0.048}^{+0.045}$ & $<10$\\ 
 \noalign{\smallskip}
$\log{U}$ & $-1.10_{-0.14}^{+0.16}$& $-1.05_{-0.12}^{+0.15}$ & $-1.01_{-0.18}^{+0.19}$ & $-1.14_{-0.16}^{+0.18}$ &  $-1.14_{-0.16}^{+0.16}$ & $-1.11_{-0.16}^{+0.19}$  & $-1.22_{-0.12}^{+0.11}$ & $2-24$\\ %
 \noalign{\smallskip}
$A_V/{\rm{mag}}$ & $0.30_{-0.08}^{+0.08}$ &   $0.36_{-0.07}^{+0.07}$ &  $0.35_{-0.08}^{+0.10}$ & $0.28_{-0.07}^{+0.10}$ & $0.29_{-0.08}^{+0.09}$ &$0.29_{-0.09}^{+0.16}$ & $0.23_{-0.05}^{+0.07}$ & $3-23$\\ %
 \noalign{\smallskip}
\hline
\noalign{\smallskip}
\multicolumn{9}{c} {s01143}   \\
 \hline
 \noalign{\smallskip}
$\log{M_*/M_{\odot}}$ & $8.89_{-0.03}^{+0.02}$ & $8.88_{-0.02}^{+0.01}$ & $8.51_{-0.11}^{+0.12}$ & $8.56_{-0.09}^{+0.09}$ & $8.53_{-0.09}^{+0.09}$   & $8.54_{-0.08}^{+0.09}$ &  $8.57_{-0.07}^{+0.10}$ & $2-58$\\
 \noalign{\smallskip}
 $\rm{SFR}/M_{\odot} \ \rm{yr^{-1}}$ & $ 8.6_{-0.6}^{+0.5}$ & $ 8.5_{-0.4}^{+0.6}$ &  $ 3.3_{-0.8}^{+1.0}$ & $3.7_{-0.6}^{+0.9}$ & $ 3.4_{-0.6}^{+0.9}$ & $3.6_{-0.7}^{+0.8}$ & $3.9_{-0.6}^{+0.9}$ & $1-62$ \\ %
 \noalign{\smallskip}
  ${\langle a \rangle}_*^{\rm{m}}/\rm{Myr}$  & $13_{-3}^{+4}$&  $10_{-0}^{+1}$ & $2.7_{-0.2}^{+0.1}$ & $3.3_{-0.4}^{+0.5}$ & $2.7_{-0.2}^{+0.3}$ & $3.1_{-0.3}^{+0.4}$   & $4.1_{-0.7}^{+0.7}$ & $23-79$ \\ %
 \noalign{\smallskip}
$Z/Z_{\odot}$  & $ 0.203_{-0.003}^{+0.005}$ &$ 0.203_{-0.003}^{+0.005}$ & $0.265_{-0.041}^{+0.047}$    & $ 0.292_{-0.057}^{+0.061}$ & $ 0.282_{-0.051}^{+0.058}$ & $0.288_{-0.056}^{+0.064}$ & $0.294_{-0.060}^{+0.069}$ & $<44$  \\
 \noalign{\smallskip}
$\log{U}$ & $-1.58_{-0.09}^{+0.06}$&  $-1.58_{-0.08}^{+0.06}$& $-0.65_{-0.23}^{+0.26}$ & $-0.86_{-0.17}^{+0.17}$ & $-0.83_{-0.19}^{+0.20}$ & $-0.88_{-0.17}^{+0.18}$ &  $-0.89_{-0.11}^{+0.18}$ & $<751$\\ %
 \noalign{\smallskip}
$A_V/{\rm{mag}}$ & $1.22_{-0.05}^{+0.05}$ &  $1.21_{-0.06}^{+0.05}$ & $1.03_{-0.13}^{+0.15}$ & $1.02_{-0.12}^{+0.12}$ &  $1.02_{-0.12}^{+0.13}$ & $1.01_{-0.12}^{+0.13}$ &  $1.01_{-0.06}^{+0.06}$ & $1-17$ \\ %
 \noalign{\smallskip}
\hline
 \noalign{\smallskip}
\multicolumn{9}{c} {s01149}   \\
 \hline
 \noalign{\smallskip}
$\log{M_*/M_{\odot}}$ & $8.87_{-0.05}^{+0.06}$ &  $8.86_{-0.05}^{+0.05}$ &$8.49_{-0.11}^{+0.08}$  & $8.64_{-0.08}^{+0.10}$  & $8.59_{-0.08}^{+0.10}$ & $8.66_{-0.10}^{+0.09}$   &  $8.70_{-0.09}^{+0.08}$ & $2-58$\\
 \noalign{\smallskip}
  $\rm{SFR}/M_{\odot} \ \rm{yr^{-1}}$ & $ 8.1_{-0.6}^{+1.0}$ &  $ 8.0_{-0.8}^{+1.0}$ & $ 3.2_{-0.7}^{+0.6}$ &  $4.7_{-0.9}^{+1.2}$ & $4.1_{-0.7}^{+1.1}$  & $4.9_{-1.1}^{+1.2}$  &  $5.4_{-1.0}^{+1.2}$ & $1-60$ \\ %
 \noalign{\smallskip}
  ${\langle a \rangle}_*^{\rm{m}}/\rm{Myr}$  & $13_{-3}^{+20}$& $11_{-1}^{+4}$& $3.4_{-0.2}^{+0.4}$ & $6.1_{-1.5}^{+2.3}$ & $4.4_{-0.7}^{+1.0}$ & $6.1_{-1.4}^{+1.9}$  & $8.9_{-2.0}^{+2.6}$ & $15-74$\\ %
 \noalign{\smallskip}
$Z/Z_{\odot}$  & $ 0.202_{-0.012}^{+0.016}$ & $ 0.202_{-0.012}^{+0.017}$ & $0.198_{-0.019}^{+0.044}$ & $ 0.188_{-0.016}^{+0.040}$ & $ 0.188_{-0.015}^{+0.040}$ &  $ 0.186_{-0.015}^{+0.046}$ & $0.184_{-0.014}^{+0.037}$ & $<9$\\ %
 \noalign{\smallskip}
$\log{U}$ & $-1.43_{-0.09}^{+0.12}$& $-1.43_{-0.09}^{+0.13}$& $-0.83_{-0.22}^{+0.26}$  & $-0.86_{-0.25}^{+0.33}$ & $-0.86_{-0.25}^{+0.32}$ &  $-0.87_{-0.26}^{+0.33}$ & $-0.85_{-0.24}^{+0.30}$ & $<298$\\ %
 \noalign{\smallskip}
 $A_V/{\rm{mag}}$ & $1.07_{-0.08}^{+0.09}$ & $1.07_{-0.08}^{+0.08}$ & $0.93_{-0.11}^{+0.11}$ &$0.94_{-0.10}^{+0.11}$  & $0.94_{-0.12}^{+0.12}$ & $0.95_{-0.12}^{+0.12}$&  $0.94_{-0.12}^{+0.11}$ & $<13$\\ %
 \noalign{\smallskip}
\hline
\end{tabular}
\end{table*}

\begin{figure*}
\centering
\textbf{Variation of galaxy parameters with adopted SFH}\par\medskip
\includegraphics[width=0.32\hsize]{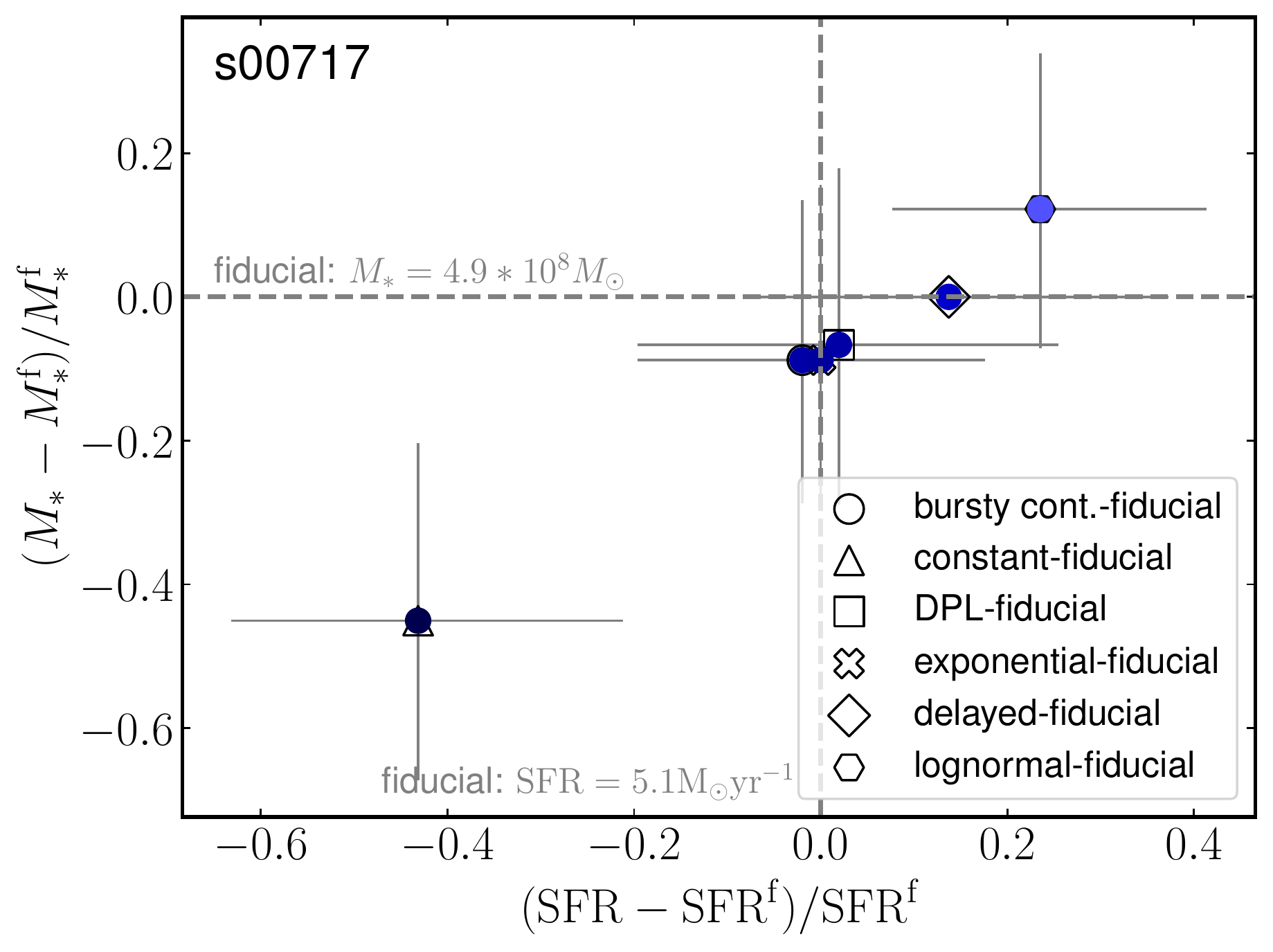}
\includegraphics[width=0.32\hsize]{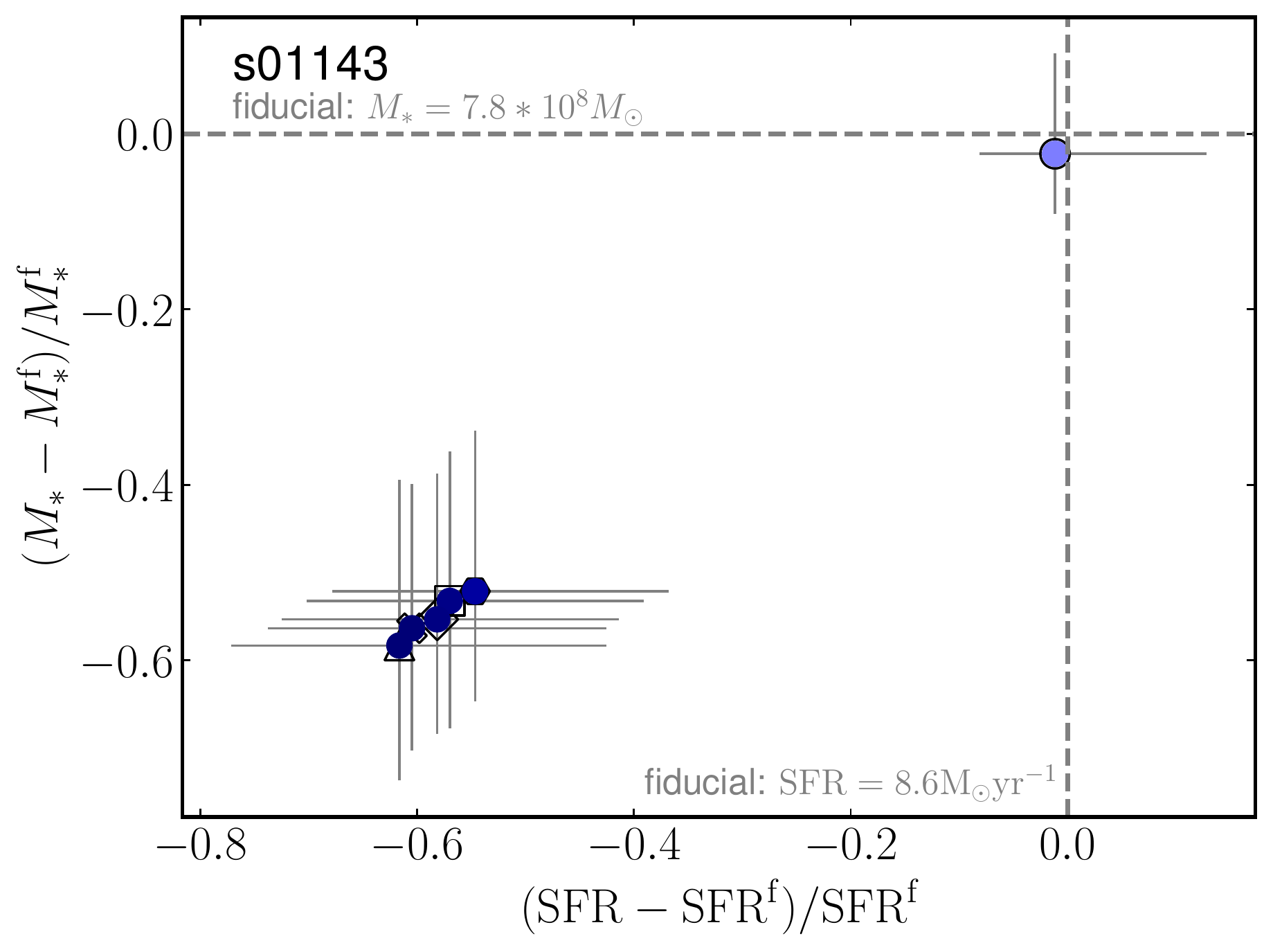}
\includegraphics[width=0.33\hsize]{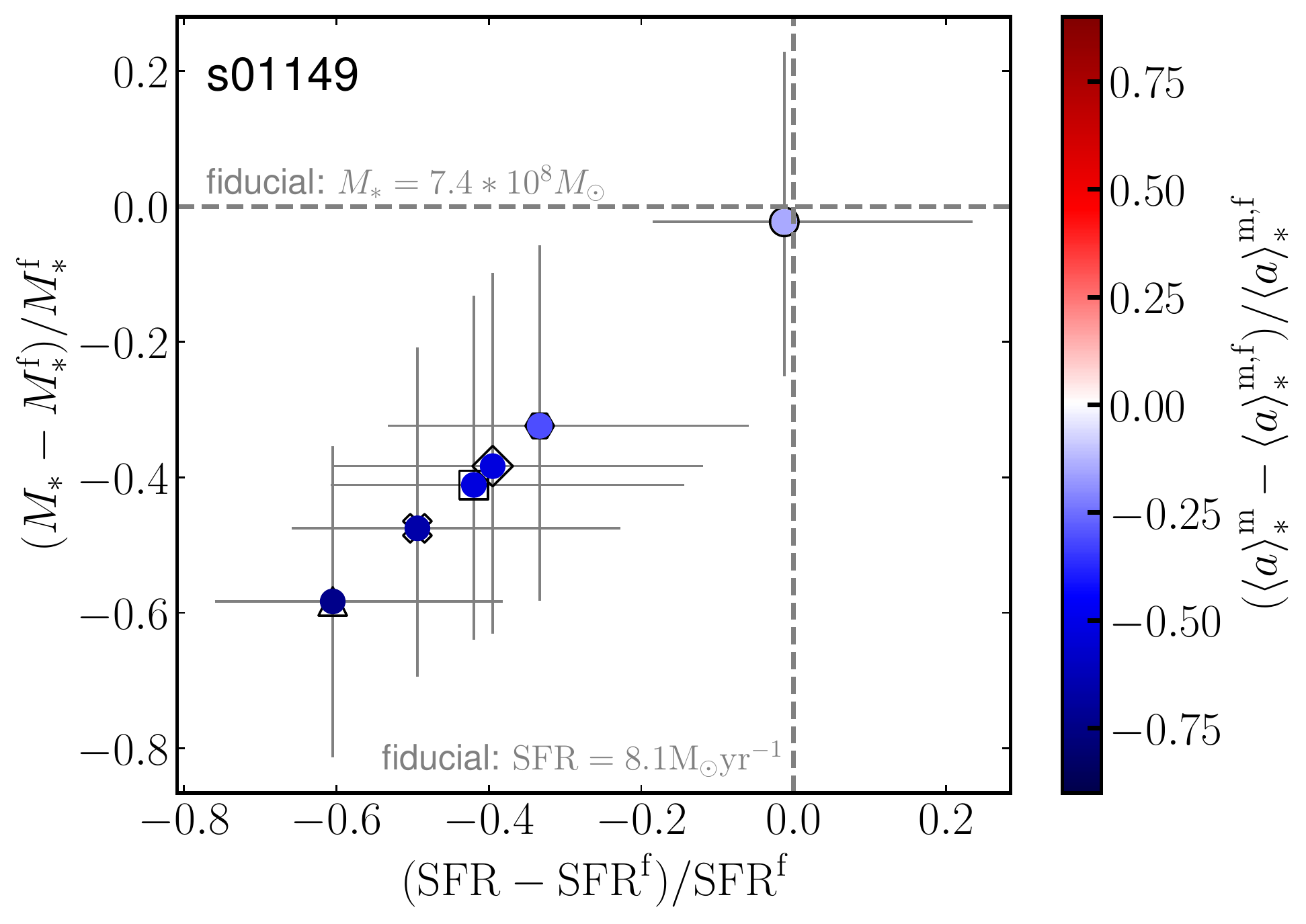}
\includegraphics[width=0.33\hsize]{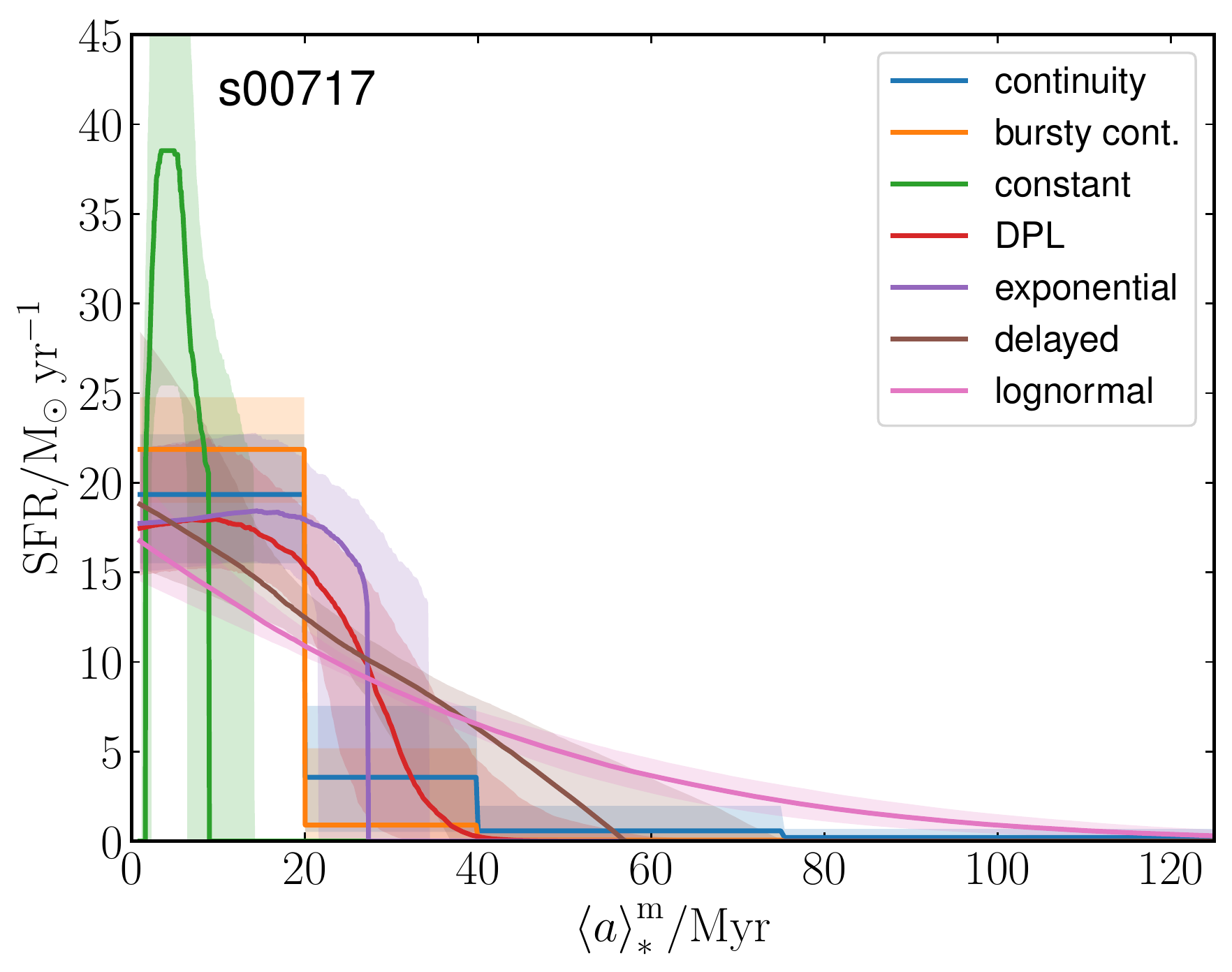}
\includegraphics[width=0.33\hsize]{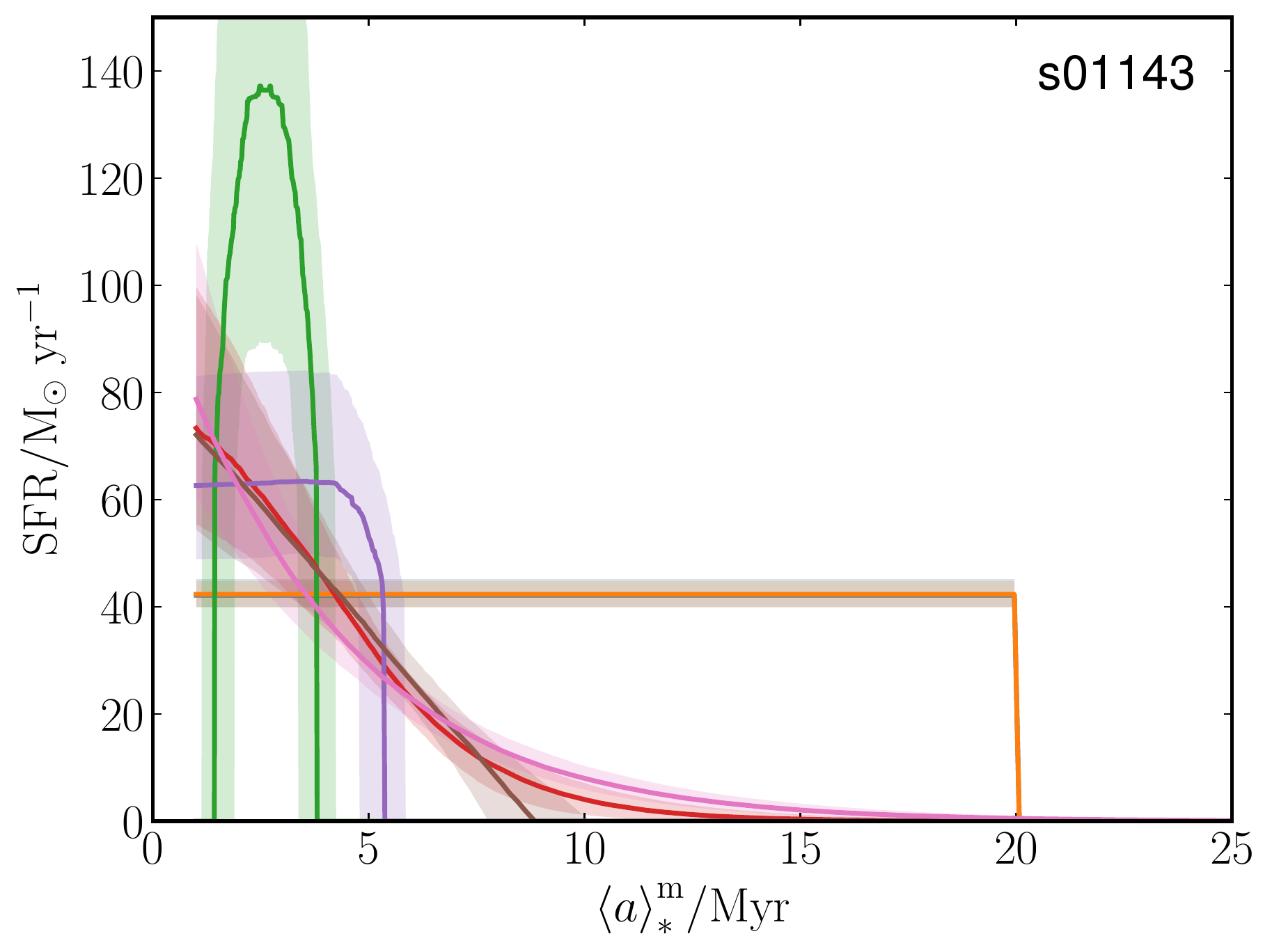}
\includegraphics[width=0.33\hsize]{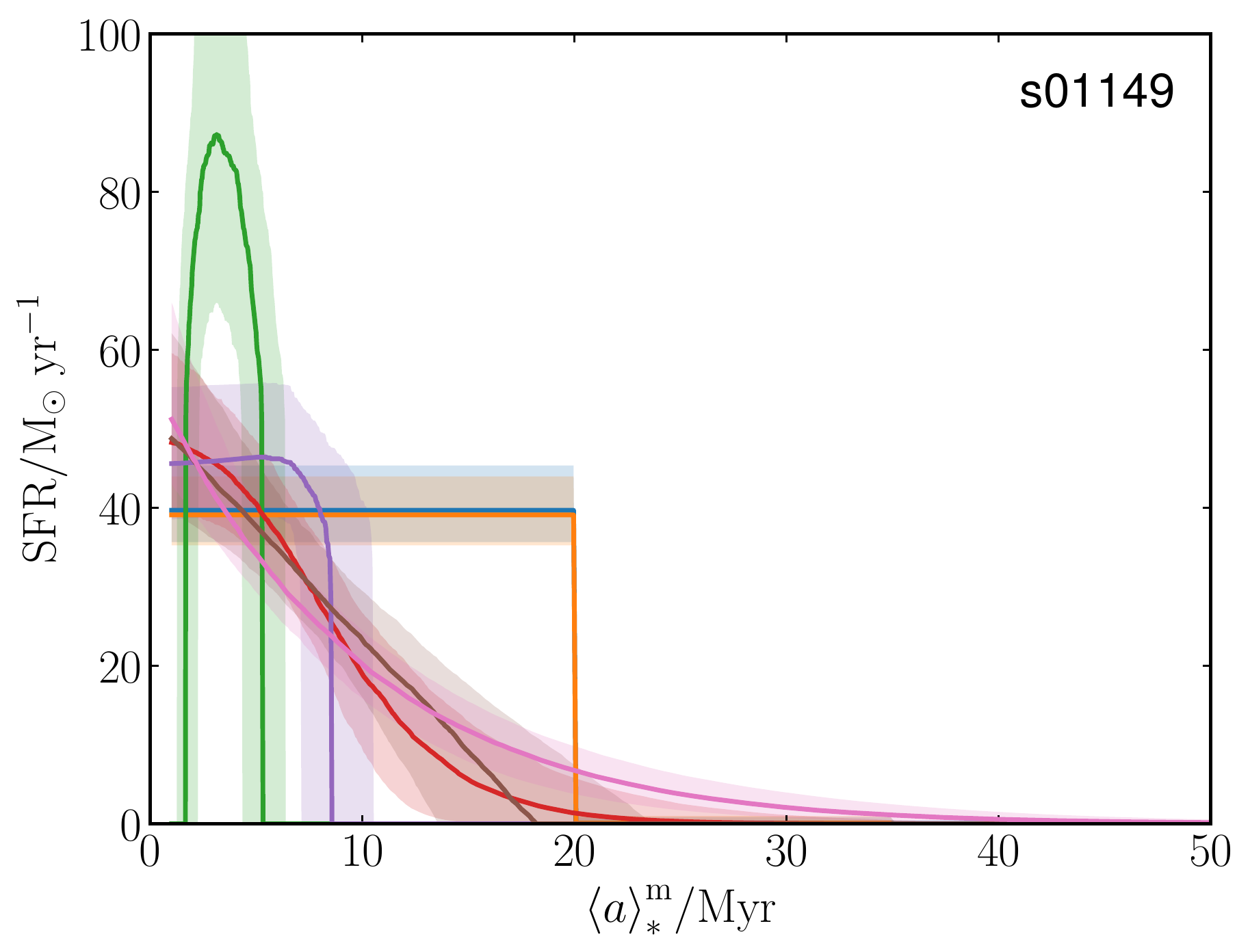}
\caption{Stellar mass ${M_*}$ and SFR, color-coded by the mass-weighted stellar age ${\langle a \rangle}_*^{\rm{m}}$ (top panels) and star formation history SFH (bottom panels), for s00717 (left), s01143 (middle), and s01149 (right). The galaxy parameters and SFHs are inferred by adopting various SFH models: the non-parametric models with continuity and bursty continuity prior, and constant, double power-law, exponential, delayed and lognormal parametric model.
\label{pars_SFH_m}
}
\end{figure*}

Most of the inferred galaxy properties vary significantly depending on the assumed SFH model (see Table \ref{table_SFH}). Particularly, the parameters closely associated to the SFH, such as the mass-weighted stellar age ${\langle a \rangle}_*^{\rm{m}}$, and to a lesser extent, the stellar mass ${M_*}$ and SFR, reveal the strongest dependence on the assumed SFH. For instance, the mass-weighted stellar age ${\langle a \rangle}_*^{\rm{m}}$ fluctuates by $>10 \sigma$ ($0.6-1.0$ dex), for the three sources. Simultaneously, the stellar mass and SFR vary by $4-5\sigma$ ($0.3-0.4$ dex) and $5-8\sigma$ ($0.3-0.4$ dex), respectively. We illustrate the fluctuations of these parameters in Fig. \ref{pars_SFH_m} (top panels).
For s01143 and s01149 galaxies, there is a systematic difference in parameters inferred from the non-parametric and parametric SFH (Fig. \ref{pars_SFH_m}, top middle and right panels). This is a consequence of the fact that the non-parametric models tend to favor a smoother SFH (Fig. \ref{pars_SFH_m}, bottom middle and right panels), and thus, longer times for mass to assemble (i.e.m higher mass-weighted ages ${\langle a \rangle}_*^{\rm{m}}$), and  thus, higher stellar masses $\log{M_*}$, and higher SFR (averaged over 100 Myr).
For the remaining source, s00717, the non-parametric SFHs are generally consistent with the more complex parametric SFH models (DPL, delayed, and lognormal), apart from the constant parametric SFH model that is biased towards a very bursty SFH (Fig. \ref{pars_SFH_m}, bottom left panel), and consequently infers shorter ${\langle a \rangle}_*^{\rm{m}}$, and lower $\log{M_*}$ and SFR (Fig. \ref{pars_SFH_m}, top left panel). 

The metallicity $Z$ shows moderate dependency on the assumed SFH model, with variations of $\sim 1-2\sigma$ ($\sim 0.05-0.15$ dex) for all the three sources. Next, the ionization parameter $\log{U}$ is largely independent (within $1 \sigma$) on the SFH for s00717, whereas it varies by $2-4\sigma$ ($\sim 0.6-0.9$ dex) for the remaining two sources, with a clear separation on the $\log{U}$ parameter estimate when using the parametric SFH with respect to the non-parametric models. Finally, the $A_V$ parameter is moderately dependent on the SFH model, with variations of up to $\sim 1\sigma$ (0.18 dex) for s00717, and up to $\sim 2-3\sigma$ ($0.05-0.08$ dex) for the remaining two galaxies; probably driven by the well-known degeneracy between the stellar age, metallicity and dust attenuation (e.g., \citealp{2001ApJ...559..620P}; \citealp{2022ApJ...927..170T}).

\begin{figure}
\centering
\includegraphics[width=0.9\hsize]{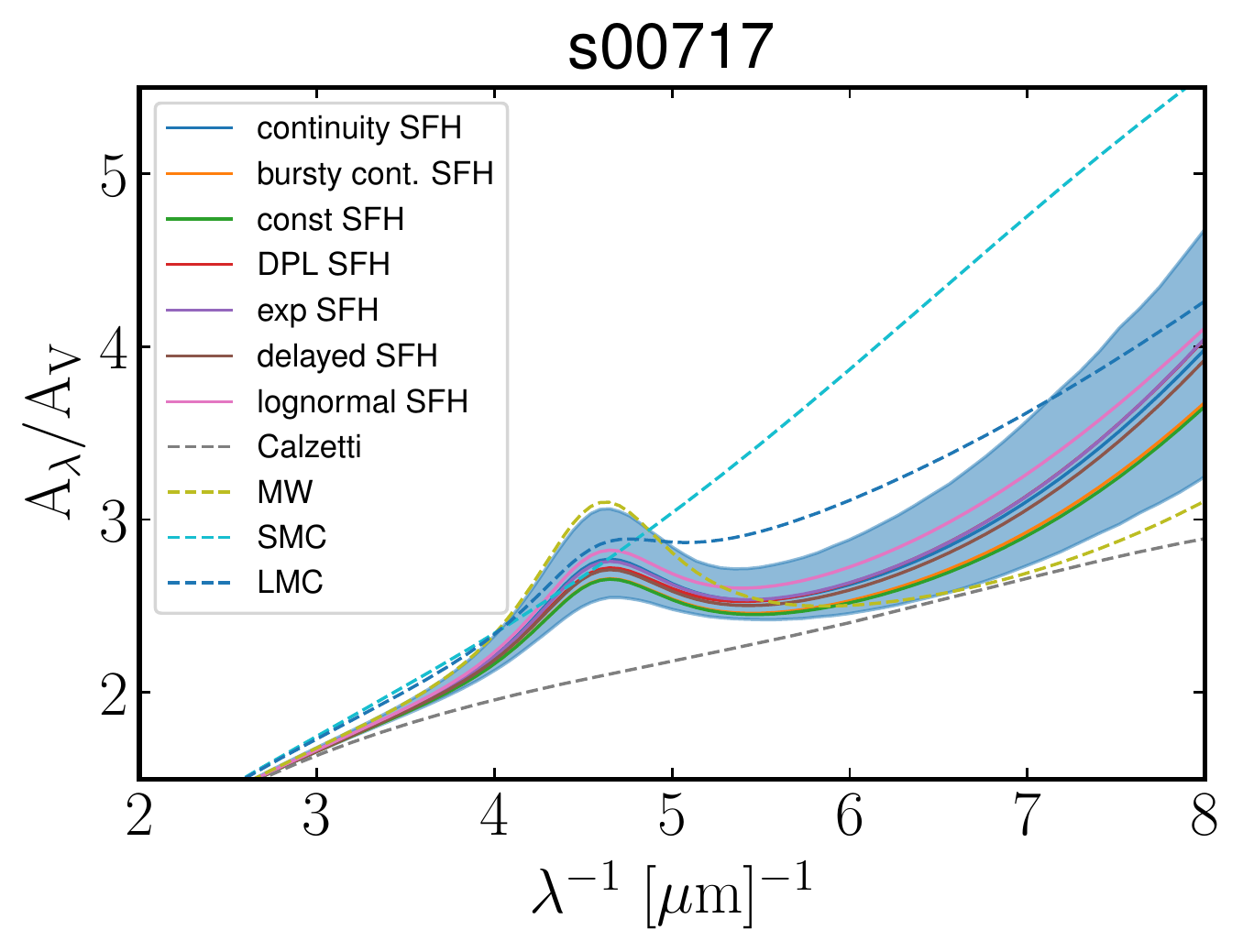}
\includegraphics[width=0.9\hsize]{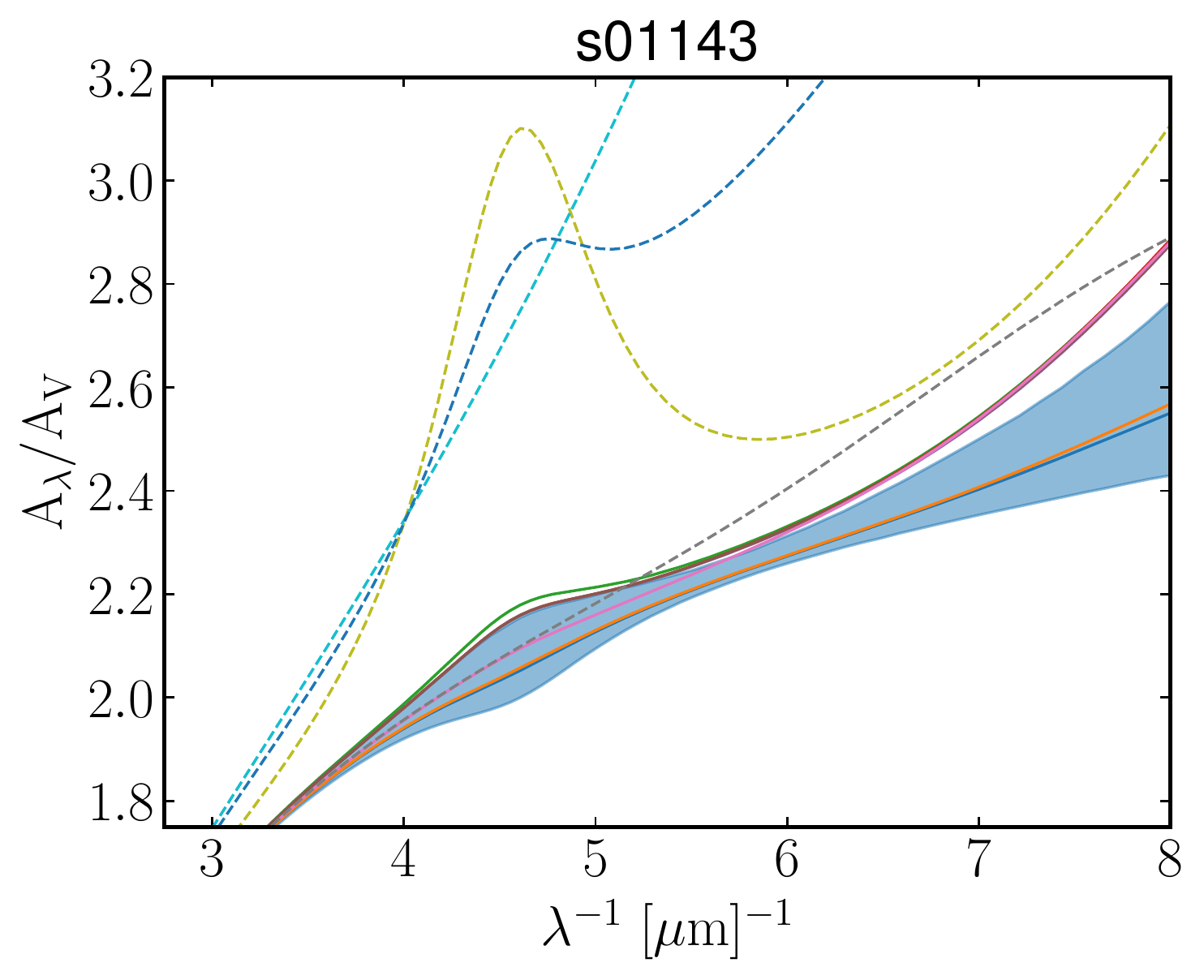}
\includegraphics[width=0.9\hsize]{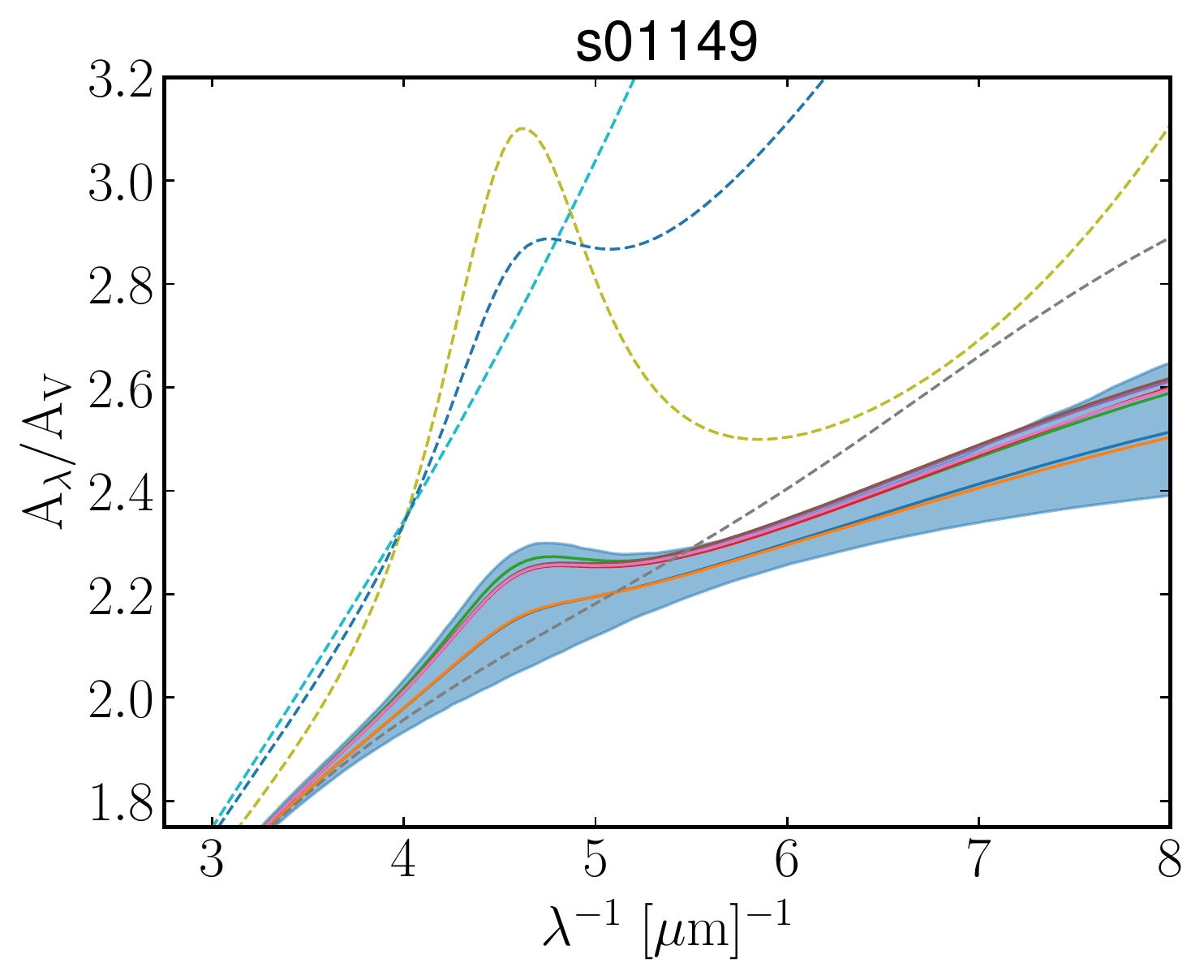}
\caption{The best-fit dust attenuation models for s00717 (top), s01143 (middle), and s01149 (bottom) derived with the SED fitting method, adopting various aforementioned SFH models (solid lines).  
$1\sigma$ uncertainties of the fiducial SFH model with the continuity prior are shown as blue shaded region. Drude model fits to conventional dust attenuation laws, frequently adopted as standard templates 
are shown for comparison (dashed lines).
\label{curves_all}
}     
\end{figure}

The dust attenuation curve is overall independent on the assumed SFH model. The properties of the dust model ($c_1-c_4$) are in the $1-1.5\sigma$ range of the fiducial model. This variation does not strongly affect the overall shape of the attenuation curve, but the amplitude of the dust absorption can fluctuate (see Fig. \ref{curves_all}), mostly driven by the variation of the $A_V$ parameter, whaich acts as the normalization of the attenuation curve.

\section{Discussion}\label{Discussion}

Within this section, we delve into the primary outcome of our research - namely, the recovered dust attenuation curves for a sample of three galaxies at $z \sim 7-8$, using our approach that builds upon the SED fitting method.
While SED fitting is an effective method for getting insight on the galaxy properties from spectra and/or photometry, its accuracy may be restricted by the uncertainties in the star formation and dust attenuation models (\citealp{2009ApJ...699..486C}; \citealp{2013MNRAS.435...87M}; \citealp{2015ApJ...806..259R}; \citealp{2019ApJ...873...44C}; \citealp{2022MNRAS.516..975T}). In this Section we compare our results on the (in)consistencies in the derived galaxy properties when adopting various dust attenuation and SFH models in the SED fitting procedure, with similar works from the literature.

\subsection{Do dust attenuation curves evolve with redshift?} \label{curve_evolution}

In Sect. \ref{curve}, we showed that the inferred dust attenuation curves of our three sources at the EoR ($z \sim 7-8$) have distinct features with respect to the standard dust templates in the local Universe. Additionally, we noted that the galaxy s00717 displays clear evidence of the $2175 \AA$ bump. On the contrary, the other two sources do not reveal any significant indications of this characteristic feature, although there are some hints of the $2175 \AA$ bump when assuming certain  parametric SFHs (Fig. \ref{curves_all}, middle and bottom panels). 

The "MW bump", a notable absorption feature in the ultraviolet spectrum, is often linked to the potential existence of carbonaceous grains and/or PAHs (\citealp{1965ApJ...142.1683S}; \citealp{2022MNRAS.514.1886S}; \citealp{2023arXiv230205468W} see \citealp{2003ARA&A..41..241D} for a review). Thus far, this characteristic feature has been identified at low-to-intermediate redshifts ($z \sim 1-3$; \citealp{refId0}; \citealp{Kriek_2013}; \citealp{Battisti_2020}; \citealp{Shivaei_2020, 2022MNRAS.514.1886S}), with only one detection reported at high-$z$ ($z \sim 6.7$; \citealp{2023arXiv230205468W}).

The characteristic $2175 \AA$ bump is typically observed in more evolved, massive, metal-rich galaxies (\citealp{refId0}; \citealp{Shivaei_2020, 2022MNRAS.514.1886S}). Although the stellar masses of the three sources of our sample are comparable, the s00717 source has a metallicity of $Z \sim 0.3 \ Z_{\odot}$ that is by a factor of $\sim 1.5$ higher, and a stellar population that is older, with the mass-weighted stellar age by a factor of $\sim 3.5$ higher, compared to the other two galaxies of our sample. The metallicity and the mass-weighted stellar age of s00717 are comparable to those of the JADES-GS-z6-0 source at $z \sim 6.7$, analyzed by \cite{2023arXiv230205468W}, which also exhibits the $2175 \AA$ absorption feature.

The strength of the $2175 \AA$ bump, relatively high metallicity $Z \sim 0.3 \ Z_\odot$ and a relatively young stellar population (${\langle a \rangle}_*^{\rm{m}} = 46 \ \rm{Myr}$) of the s00717 galaxy at $z\sim 6.9$, suggests the need for a dust production mechanism on short timescales, that is possible to occur in the SNeII remnants (\citealp{2015MNRAS.451L..70M}; \citealp{2019A&A...624L..13L}; \citealp{2023arXiv230205468W}) or dust grain growth in the ISM (\citealp{2013EP&S...65..213A}; \citealp{2019A&A...624L..13L}).

\subsection{Do galaxy properties depend on the dust model?} \label{dust_discussion}

In Sect. \ref{comparison}, we have shown that {\it a priori} adopting a standard dust attenuation curve when performing the SED fitting can lead to large uncertainties in the estimates of the important global physical properties such as $\log{M_*}$ and SFR, and dust attenuation parameter $A_V$ of the source (by up to $\sim 0.35$ dex, $\sim 0.34$ dex, and $\sim 0.38$ dex, respectively), if the assumed attenuation curve significantly departs from the fiducial attenuation curve.
This is in overall agreement with a similar works from the literature (\citealp{2015ApJ...806..259R}; \citealp{2016ApJS..227....2S}; \citealp{2022MNRAS.516..975T}).
For instance, \cite{2015ApJ...806..259R} found that the stellar masses and SFR differ by up to $\sim 0.16$ dex and $\sim 0.30$ dex, respectively by changing the dust attenuation law; they compared the Calzetti and the SMC curve, with respect to the attenuation curve derived in their work, and adopted the exponentially rising SFH as their fiducial model, for a sample of 224 star-forming galaxies at $z \sim 1.4-2.6$.
Next, \cite{2016ApJS..227....2S} found that a fraction of galaxies from their sample of $\sim 700\  000$ galaxies at $z < 0.3$, has stellar masses that vary by up to 0.4 dex, with the adopted dust law; they compared the Calzetti and their modified dust law, and assumed the two-component exponential and delayed SFHs, with \texttt{CIGALE} (\citealp{2009A&A...507.1793N}).
Finally, \cite{2022MNRAS.516..975T} found a weaker dependence on the stellar mass ($\lesssim 0.09$ dex) and a more significant dependence on the SFR ($\sim 0.4$ dex), on the choice of the dust law. \cite{2022MNRAS.516..975T} considered the Calzetti, SMC and MW curves, assuming a constant parametric SFH as fiducial, with the \texttt{BEAGLE} code (\citealp{2016MNRAS.462.1415C}).

\subsection{Do galaxy properties depend on the SFH model?} \label{Discuss_curves}

In Sect. \ref{SFH} we have explored if the inferred galaxy properties depend on the assumed SFH model. Our results indicate that significant inconsistencies emerge in the inferred properties related to the SFH, when adopting different SFH model. For instance, the mass-weighted stellar age can vary by up to an order of magnitude ($0.6-1.0$ dex), whereas the dependence is weaker for the stellar mass and SFR ($0.3-0.4$ dex).

\subsubsection{Parametric SFH models}

\cite{2018MNRAS.480.4379C} tested \texttt{BAGPIPES} using mock photometric data of massive ($\log{M_*/M_{\odot}} > 10$) quenched galaxies in the redshift range of $0.5 < z < 2.5$. They compared the parametric SFH models: the double power-law (DPL) and exponentially declining (exponential) SFH model. The study found that the DPL SFH model recovered unbiased properties of the simulated sources, while the exponential SFH model led to a slight overestimate of the stellar mass by $\sim 0.06$ dex or $\sim 15\%$. 

In an independent analysis on a sample of $z \sim 0$ mock galaxy photometric data, \cite{2019ApJ...873...44C} found that the stellar mass, SFR and mass-weighted age vary by at least $\sim 0.1$, $0.3$, and $0.2$ dex, respectively, with the assumption on the four parametric SFH model: the DPL, exponential, delayed exponentially declining (delayed), and lognormal. However, our results indicate variations of up to 0.11, 0.12, and 0.36 dex, for the three parameters, respectively, taking into account the four afore-mentioned models. 

Finally, \cite{2019ApJ...873...44C} shown that recovering the parameters depends more on the shape of the true SFHs than on the choice of the SFH model itself. Therefore, they concluded that there is no one universal (parametric) SFH model that is able to provide the best fit on the true SFH and recover the unbiased properties of all the individual sources at all redshifts. 
 
The exponentially declining model, often used as a fiducial model for nearby sources, generally does not provide a good fit for rising SFH of intermediate-to-high redshift sources ($z>2$; e.g., \citealp{2011ApJ...738..106W}; \citealp{Reddy_2012}). The double power-law can be a good choice for fitting the massive intermediate-$z$ quiescent galaxies with older stellar population (\citealp{2018MNRAS.480.4379C}). The constant and delayed SFH models often provide a good fit to the SED of high-$z$ ($z > 6$) sources (\citealp{2021MNRAS.505.3336L}; \citealp{2021MNRAS.501.1568F}). Particularly, the constant SFH model appears to be a more adequate choce for fitting the SEDs with the characteristic "Balmer jump", produced by very young stellar populations ($a_* < 10 \ \rm{Myr}$) in low-mass, star-forming galaxies at high-$z$ (\citealp{2023MNRAS.518L..45C}). Finally, some theoretical works found that exponentially rising SFHs are favored (\citealp{2017MNRAS.471.4128P, 2022MNRAS.513.5621P}).

\subsubsection{Non-parametric SFH models}

Parametric SFH models are broadly used, but are often not flexible enough to fit the "true" galaxy SFHs in all their complexity (\citealp{2019ApJ...876....3L}; \citealp{2020ApJ...904...33L}). A favorable alternative are more flexible, non-parametric SFH models which are able to fit complex SFHs with greater variations in SFR (\citealp{2019ApJ...876....3L}), often measured in high-$z$ ($z > 6$) galaxies (\citealp{2022MNRAS.516..975T};  \citealp{2022arXiv220605315W}). One of the drawbacks of the non-parametric models is that they tend to be more computationally expensive with respect to the parametric models, since typically they yield a larger parameter space to explore by the Bayesian sampler (e.g., \citealp{2019ApJ...876....3L}).
Another drawback is that the flexibility of non-parametric SFH models is somewhat limited by the assumed priors, which either weight in favor of a smooth or chaotic SFH (\citealp{2019ApJ...876....3L}; \citealp{2022arXiv220605315W}; \citealp{2022ApJ...927..170T}).
Therefore, when assuming the non-parametric SFH models, it is important to choose a well-motivated prior that best mimics the realistic SFH, as it can have a significant impact on the SED fitting and the inferred galaxy properties (\citealp{2019ApJ...876....3L}; \citealp{2022arXiv220605315W}; \citealp{2022ApJ...927..170T}; \citealp{2022MNRAS.516..975T}). However, the flexible non-parametric SFHs still provide a less biased fit with more accurate error estimates with respect to the more rigid parametric SFH models (\citealp{2019ApJ...876....3L}; \citealp{2020ApJ...904...33L}). 

Overall, the non-parametric SFH models, and particularly models that explicitly weight against sharp transitions in the SFR history (for instance, models with the continuity prior) typically lead to more extended and complex SFHs, and fit older stellar populations and thus, recover higher ${\langle a \rangle}_*^{\rm{m}}$. Extended SFHs allow longer stellar mass assembly, and consequently, significantly larger inferred stellar masses (Fig. \ref{pars_SFH_m}; see also e.g., \cite{2022MNRAS.516..975T}; \citealp{2022arXiv220605315W}).
On the contrary, the more rigid parametric SFH models are not flexible enough to fit both the recent burst of SFR originating from a young stellar population and a potentially more dominant older stellar population at earlier times. Consequently, the inferred stellar masses, SFRs and ages of these sources will be systematically lower than their fiducial values. This is particularly evident for star-forming galaxies at the EoR, where a very young stellar population ($< 10 \rm{Myr}$) can outshine the possibly more massive, but fainter older stellar population (the so-called "outshining" effect; \citealp{2019ApJ...876....3L}; \citealp{2022arXiv220605315W}; \citealp{2022MNRAS.516..975T}; \citealp{2022ApJ...927..170T}.
For instance, \cite{2022arXiv220605315W} and \cite{2022MNRAS.516..975T} found that adopting the non-parametric SFH model (with continuity prior) led to inferred stellar masses that were approximately an order of magnitude larger ($\gtrsim 1 \ \rm{dex}$) than those obtained using the parametric constant SFH model. However, in our study, we observed smaller fluctuations in stellar mass of around $\sim 0.4 \ \rm{dex}$ when comparing the same SFH models.

\section{Summary and conclusions} \label{Conclusions}
 
In this paper we presented a new method based on the \texttt{BAGPIPES} SED fitting code and an analytical dust attenuation law model which allows to simultaneously constrain the fundamental physical properties of the galaxy, along with the dust attenuation curve.
The flexibility of this dust attenuation parameterization allows to fit the SEDs of galaxies reddened by any of the well-known (e.g., the Calzetti, the SMC, and the MW), as well as any unconventional dust attenuation curve. We test our tool against synthetic spectra, attenuated by conventional attenuation templates. We apply our tool on Near Infrared Spectrograph (NIRSpec) {\it JWST} spectroscopic observations of a sample of three star-forming galaxies at the Epoch of Reionization. Our main results are: 

\begin{itemize}
    \item We performed the SED fitting of the {\it JWST} spectroscopic data and constrained the global properties along with the dust attenuation curves of our three galaxies at $z \sim 7-8$.
    The shape of the constrained dust attenuation curve of the s00717 galaxy resembles that of the LMC and MW curve, whereas the shape of the attenuation curve of the remaining two sources, s01143 and s01149 are similar to the Calzetti curve, all the three curves show distinct features at lower wavelengths.
    We report the presence of the $2175 \AA$ bump in one out of three galaxies of our sample. This feature, frequently associated with the existence of the small carbonaceous dust grains and/or PAHs, is one of the first indications of the presence of the PAHs in a galaxy at the EoR.
    \item We investigated if the constrained physical properties change with {\it a priori} assumption of one of the standard attenuation templates in the SED fitting procedure.
    We showed that an important bias is introduced in the estimates of some of the fundamental physical properties of galaxies ($\log{M_*}$ and SFR, and dust attenuation parameter $A_V$), with fluctuations of up to $\sim 0.4$ dex, if the adopted attenuation curve differs from the fiducial dust attenuation curve.
    \item We probed the sensitivity of the fundamental properties of our galaxies on the assumed SFH model.
    We found that some properties related to the SFH can change significantly, with the mass-weighted age changing by an order of magnitude (1 dex), whereas the stellar mass and SFR vary by up to 0.4 dex. On the contrary, the inferred dust attenuation curve properties are independent on the assumed SFH model. 
\end{itemize}
In the near future, applying our new robust method on the {\it JWST} spectroscopic and photometric data will provide the means to characterize the dust attenuation curves, along with the fundamental properties of a large sample of galaxies at the EoR. This will greatly enhance our understanding of the dust distribution, composition, and production mechanisms in the earliest and most distant galaxies in the Universe.

\begin{acknowledgements}
       VM would like to thank Adam Carnall for his help and useful comments regarding the \texttt{BAGPIPES} code.
       VM would like to thank James Davies and Samuel Gagnon-Hartman for their useful comments on the Bayesian Analysis.
       VM, AF, and AP acknowledge support from the ERC Advanced Grant INTERSTELLAR H2020/740120. Partial support (AF) from the Carl Friedrich von Siemens-Forschungspreis der Alexander von Humboldt-Stiftung Research Award is kindly acknowledged.
       SC acknowledges support by European Union's HE ERC Starting Grant No. 101040227 - WINGS
       Any dissemination of results must indicate that it reflects only the author’s view and that the Commission is not responsible for any use that may be made of the information it contains.
       We gratefully acknowledge computational resources of the Center for High Performance Computing (CHPC) at SNS.
\end{acknowledgements}

%
%
\bibliographystyle{aa} 
\bibliography{biblio} 

\begin{appendix}


\section{Fitting the synthetic spectra} \label{fit_syn}

We use BAGPIPES and our method outlined in Sect. \ref{method}, to generate the synthetic spectra. We fix the global, SFH, dust attenuation and nebular emission parameters of the model galaxy to be identical to the inferred parameters of our sample galaxies (Table \ref{params_gals}). The wavelength range and grid of the synthetic spectra are set to be identical to the wavelength range and grid of the observed spectra of the three sources. The flux density uncertainties are taken directly from the observed spectra. Synthetic spectra and its uncertainties are then masked below the Lyman break, in the same manner as for the observed spectra, due to the significant noise at lower wavelenghts. 

Next, we fit the synthetic spectra by the very same model, allowing the free properties of the synthetic galaxy to vary in the range as outlined in Table \ref{params}, when assuming the non-parametric SFH models. In case of adopting one of the parametric SFH model, the additional free parameters of specific models are allowed to fluctuate in the limits summarized in Table \ref{params_limits}. We test and validate our method by fitting the synthetic spectra, and recovering the initial properties of the model galaxies, SFH model and the adopted standard dust attenuation curve. 

\begin{figure*}
\centering
\includegraphics[width=0.56\hsize]{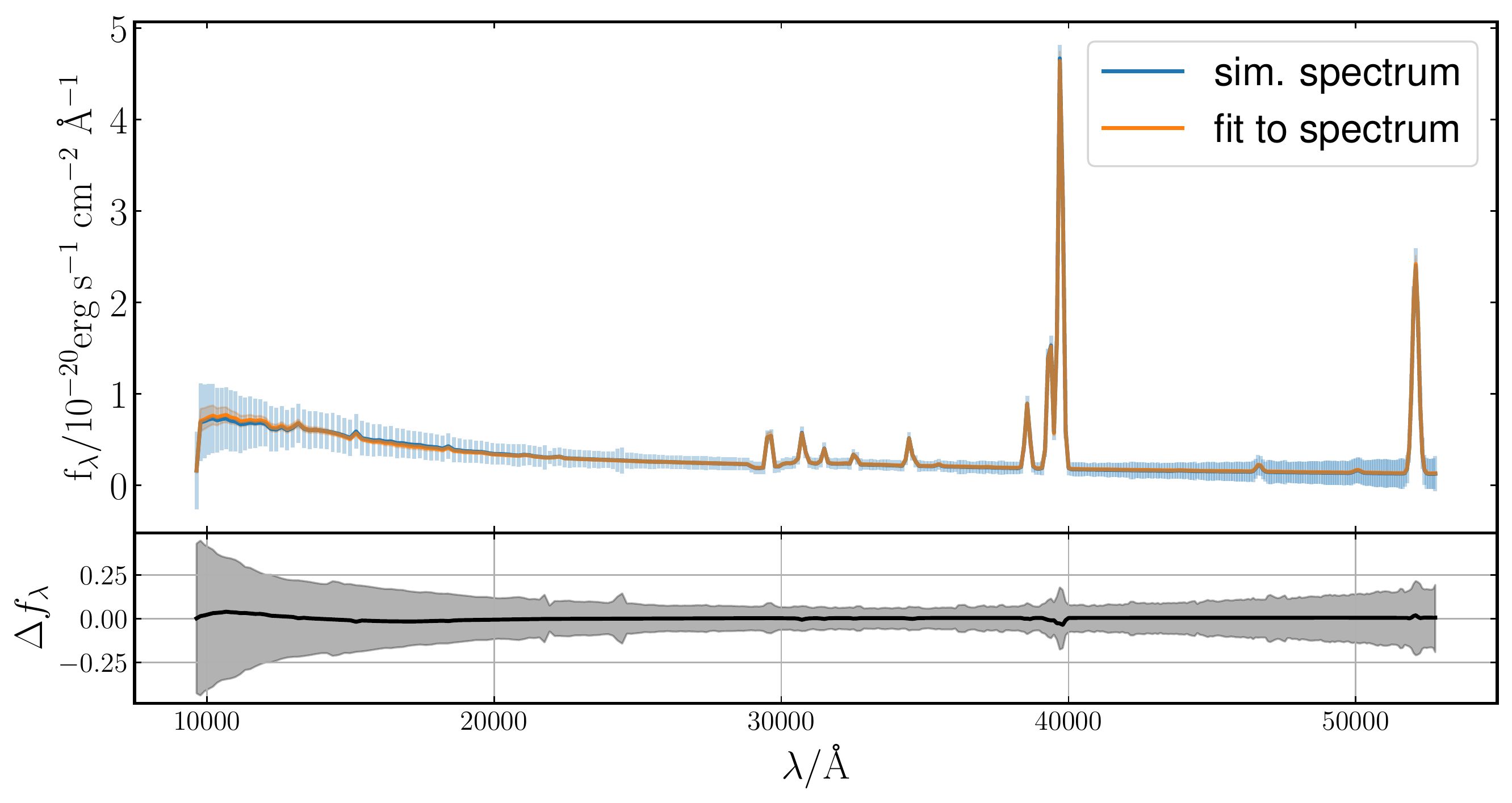}
\includegraphics[width=0.38\hsize]{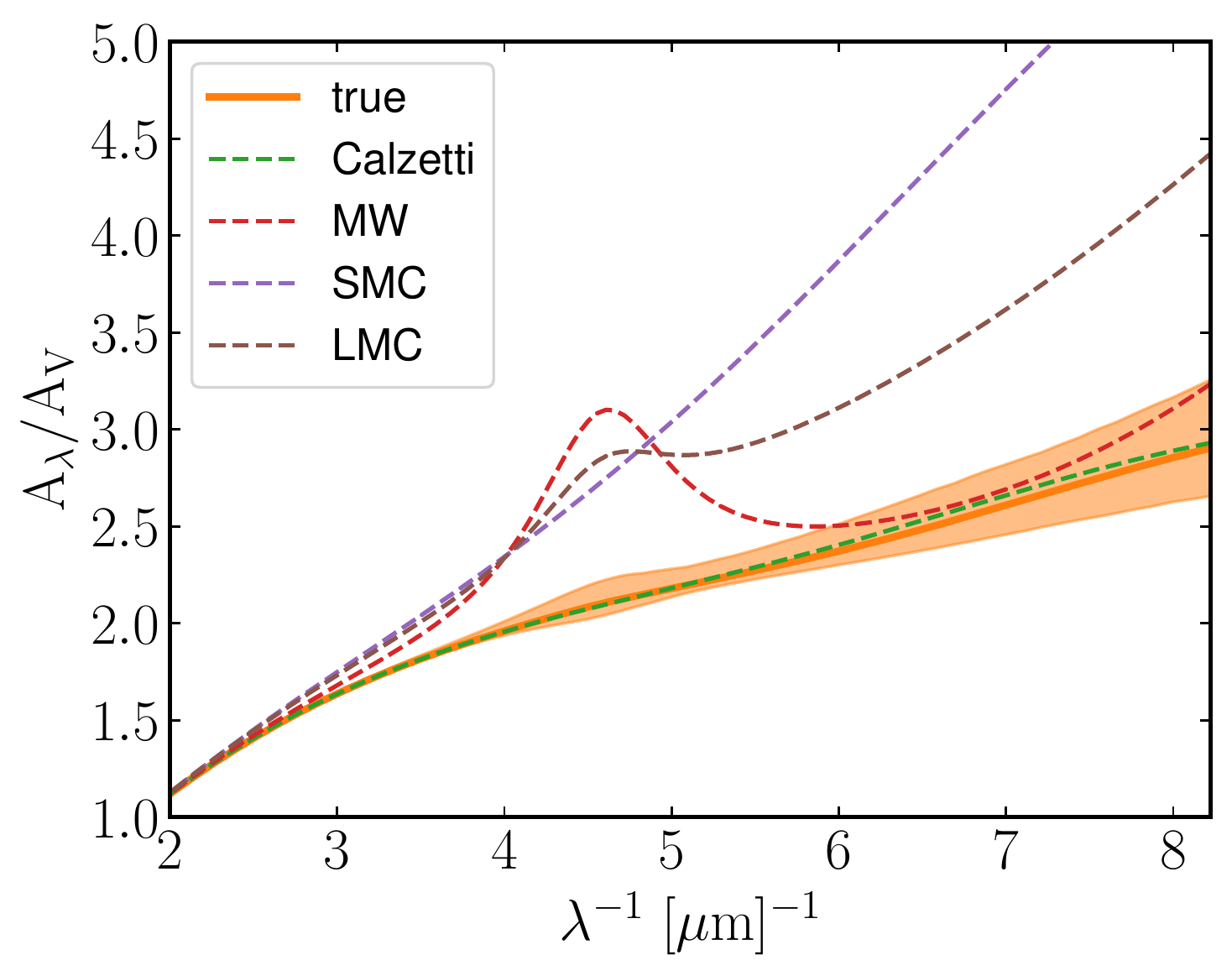}
\includegraphics[width=0.56\hsize]{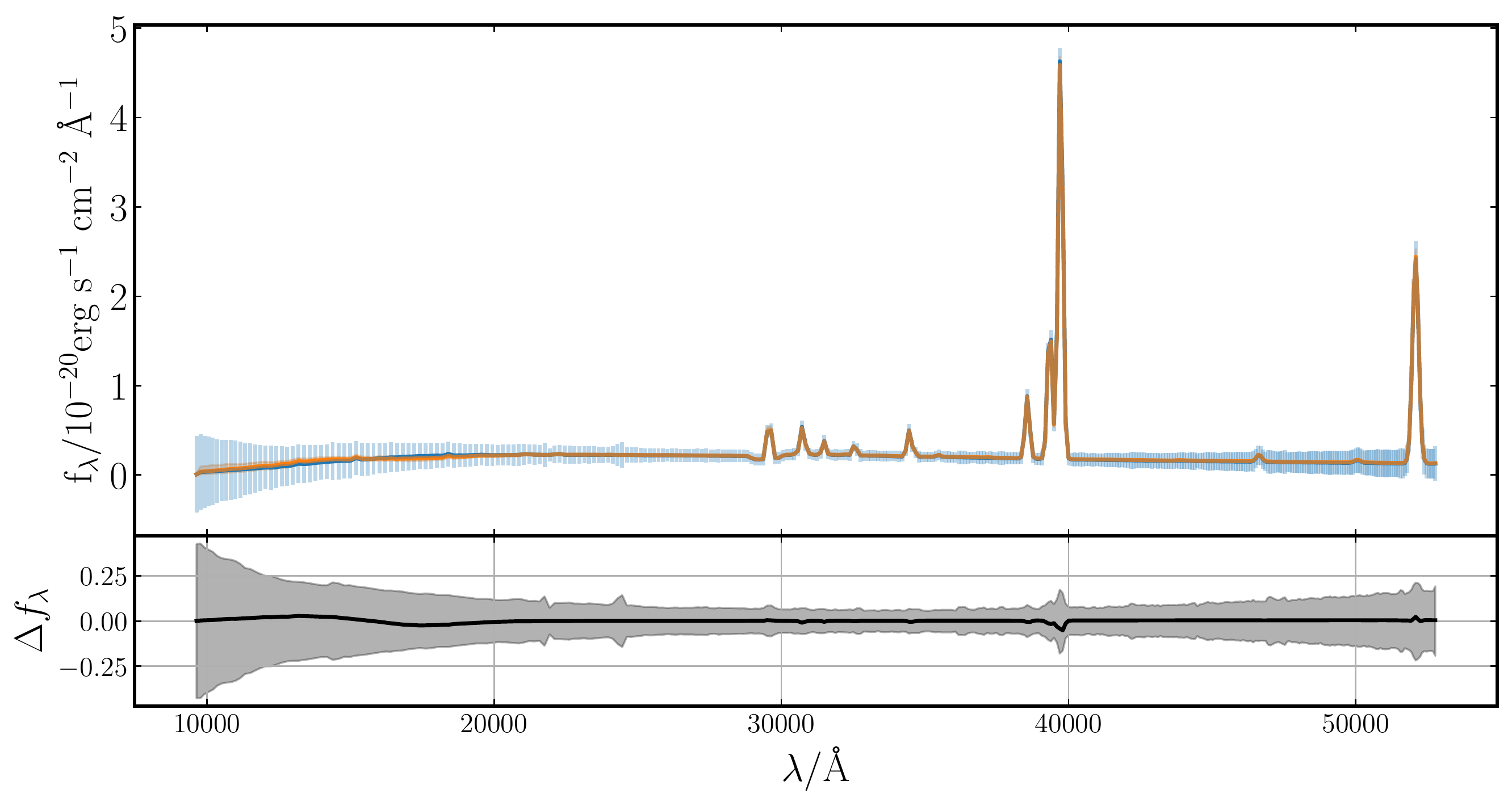}
\includegraphics[width=0.38\hsize]{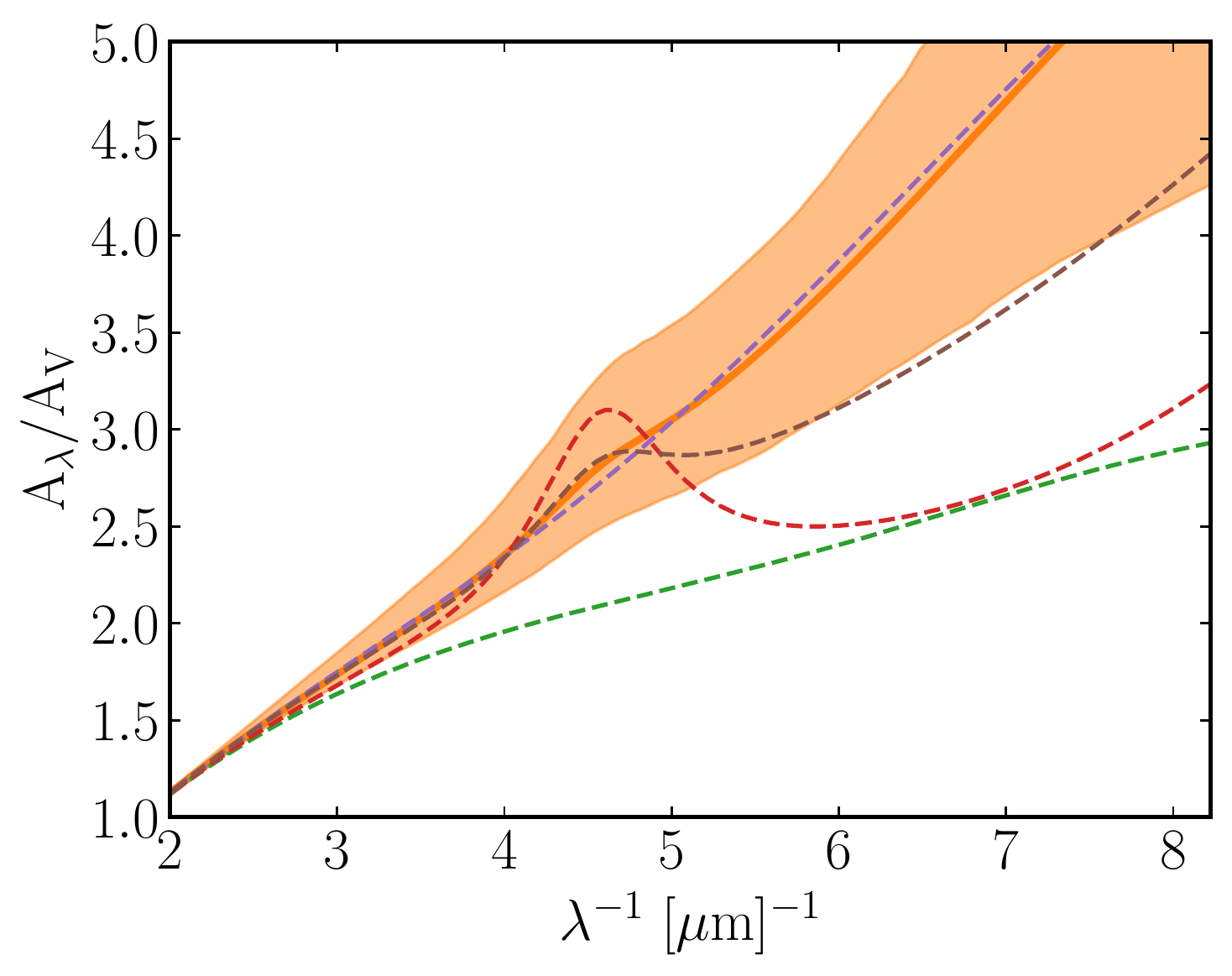}
\includegraphics[width=0.56\hsize]{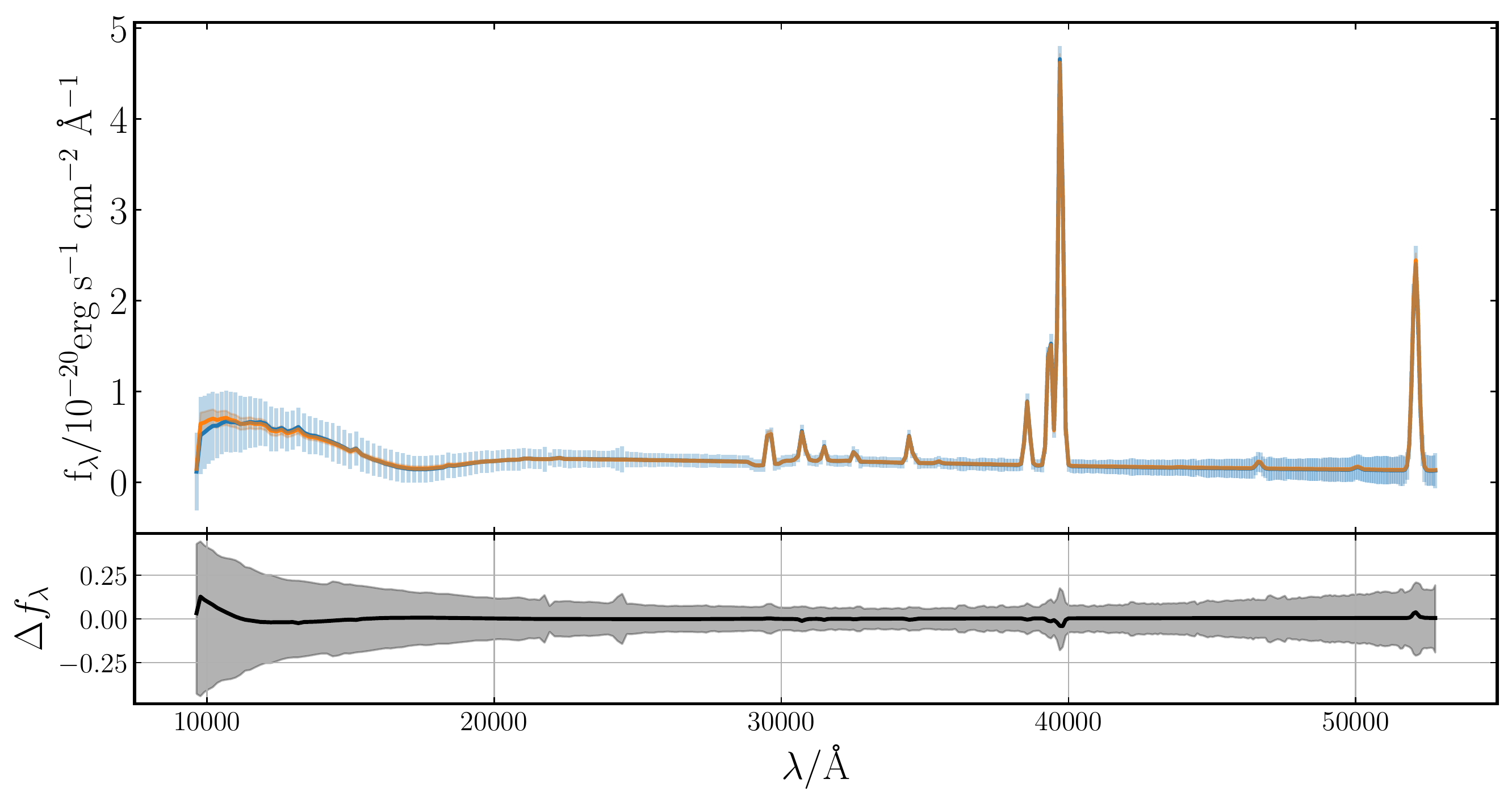}
\includegraphics[width=0.38\hsize]{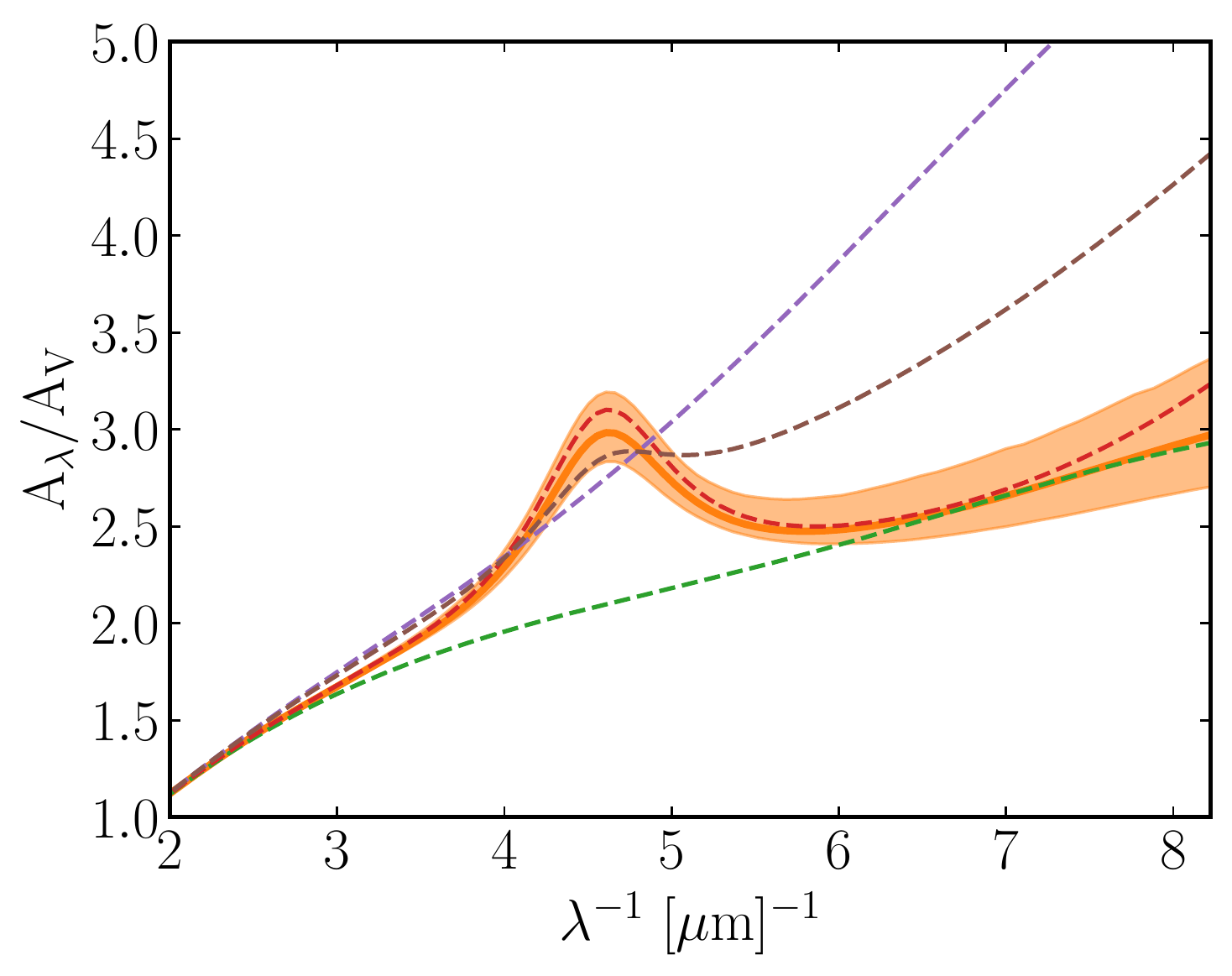}
 \caption{Top left: Synthetic dust-attenuated spectra of s01143 source, generated by using the SED fitting method, assuming the fiducial SFH, and adopting one of the standard dust curves: Calzetti, SMC, and MW (from top to bottom, respectively). The model spectra of our source are shown in blue, with flux uncertainties illustrated in pale blue. Orange and pale orange color indicate the best-fit posterior spectra with 1$\sigma$ uncertainties, respectively. Bottom left: The residuals of the best-fit on the synthetic spectra $\Delta f_{\lambda}$ with 1$\sigma$ uncertainties. Right: The corresponding best-fit dust attenuation model with 1$\sigma$ uncertainties, on conventional dust templates: Calzetti, SMC, and MW (from top to bottom). Drude model fits to dust curves: the Calzetti, the SMC, the LMC, and the MW curve are shown as green, red, purple, and brown dashed lines, respectively.
 \label{att_spectra}
 }
\end{figure*}

We showcase the usage of our method on example synthetic spectra (Fig. \ref{att_spectra}, blue spectra in left panels), generated by fixing all the parameters of the model source to be equal to the constrained properties of the s01143 galaxy (Table \ref{params_gals}), except the dust attenuation properties ($c_1-c_4$), which are set to the values of one of the conventional dust curves: Calzetti, SMC, and MW (Table \ref{table_pars}).  Next, we fit the synthetic spectra of the s01143 source, and demonstrate that we can recover any of the {\it a priori} adopted, standard dust attenuation templates (Fig. \ref{att_spectra}, left and right panels, respectively).

\begin{figure*}
\centering
\includegraphics[width=\hsize]{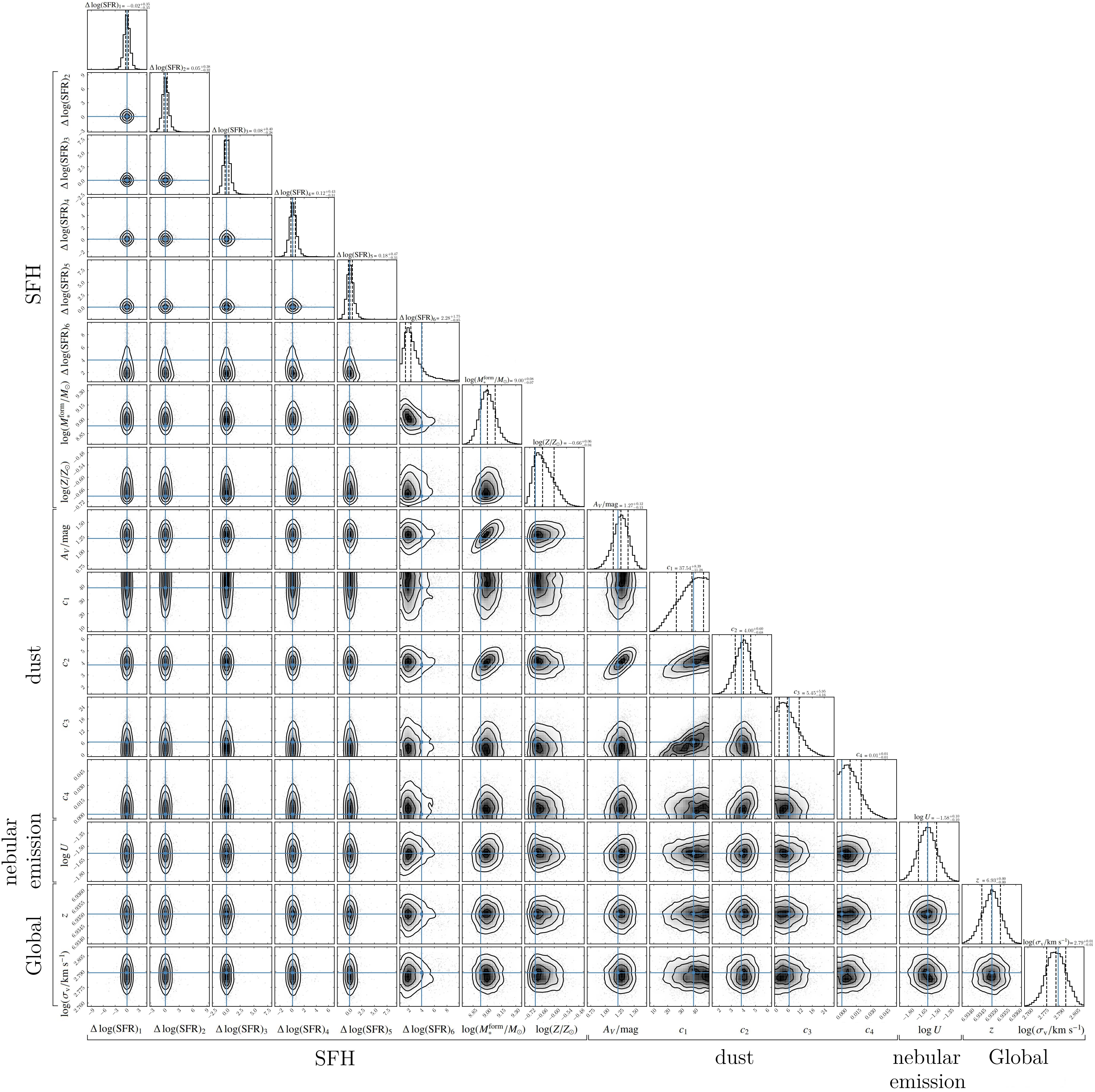}
 \caption{Corner plot illustrating the 1D and 2D projections of the posterior distribution in parameter space obtained from the SED fitting on the synthetic spectrum generated by fixing the model parameters to those of the s01143 source, assuming the SMC curve, and a fiducial SFH model. The initial (true) properties of the model galaxy are shown in blue.
 \label{corner_syn}
 }    
\end{figure*}

In Fig. \ref{corner_syn} we show a corner plot of the posterior of all the physical properties of the model galaxy, including the properties of the adopted dust attenuation curves. The initial properties of the model galaxy are within the $1-2\sigma$ uncertainties of the posterior distribution (Fig. \ref{corner_syn}, blue and black, respectively). 

\section{Corner plots of s01143 and s01149}

\begin{figure*}
\centering
\includegraphics[width=\hsize]{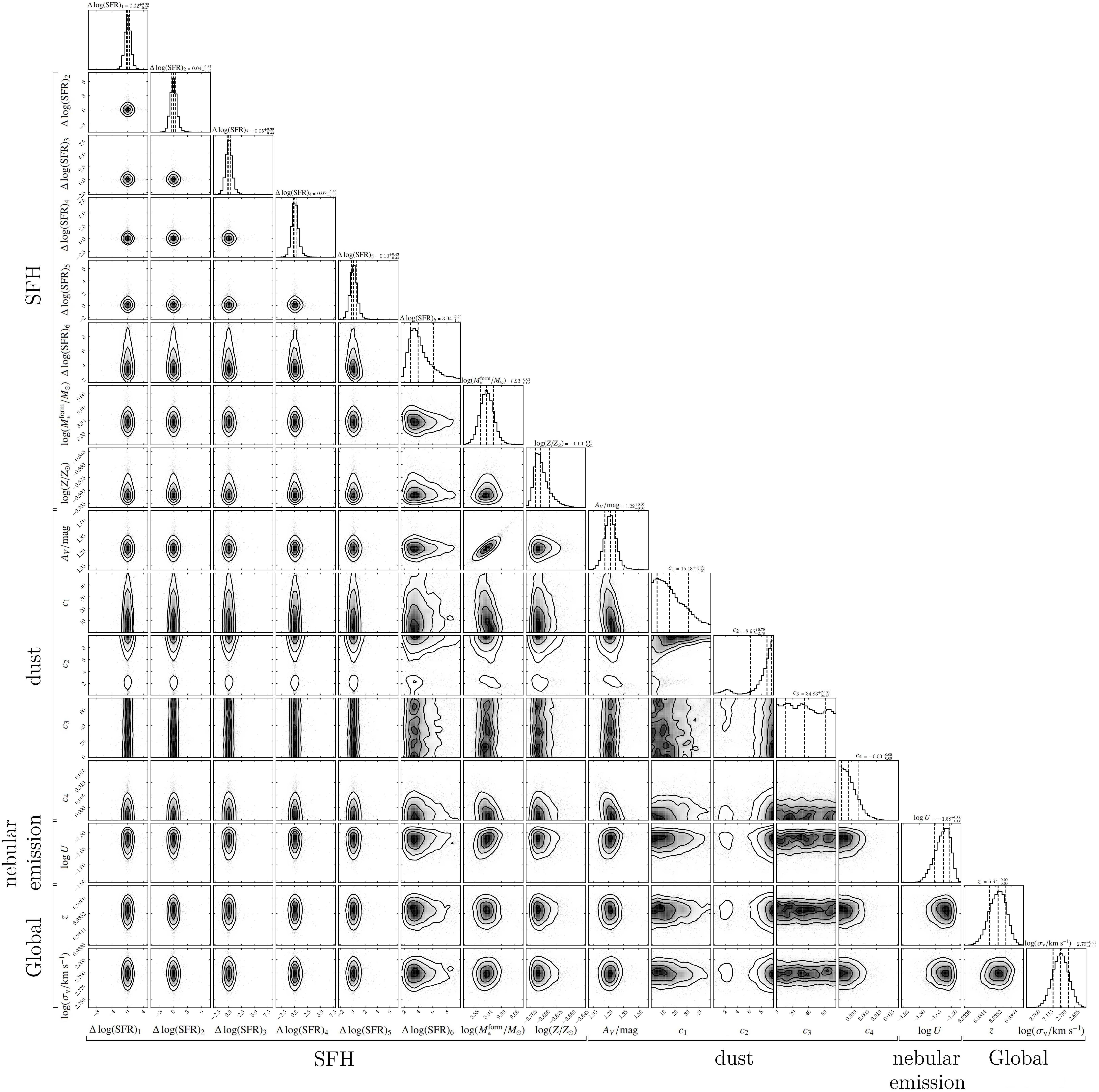}
 \caption{Corner plot illustrating the the 1D and 2D projections of the posterior distribution of the parameters derived from the SED fitting on spectrum of s01143.
 \label{corner2}
 }
\end{figure*}

\begin{figure*}
\centering
\includegraphics[width=\hsize]{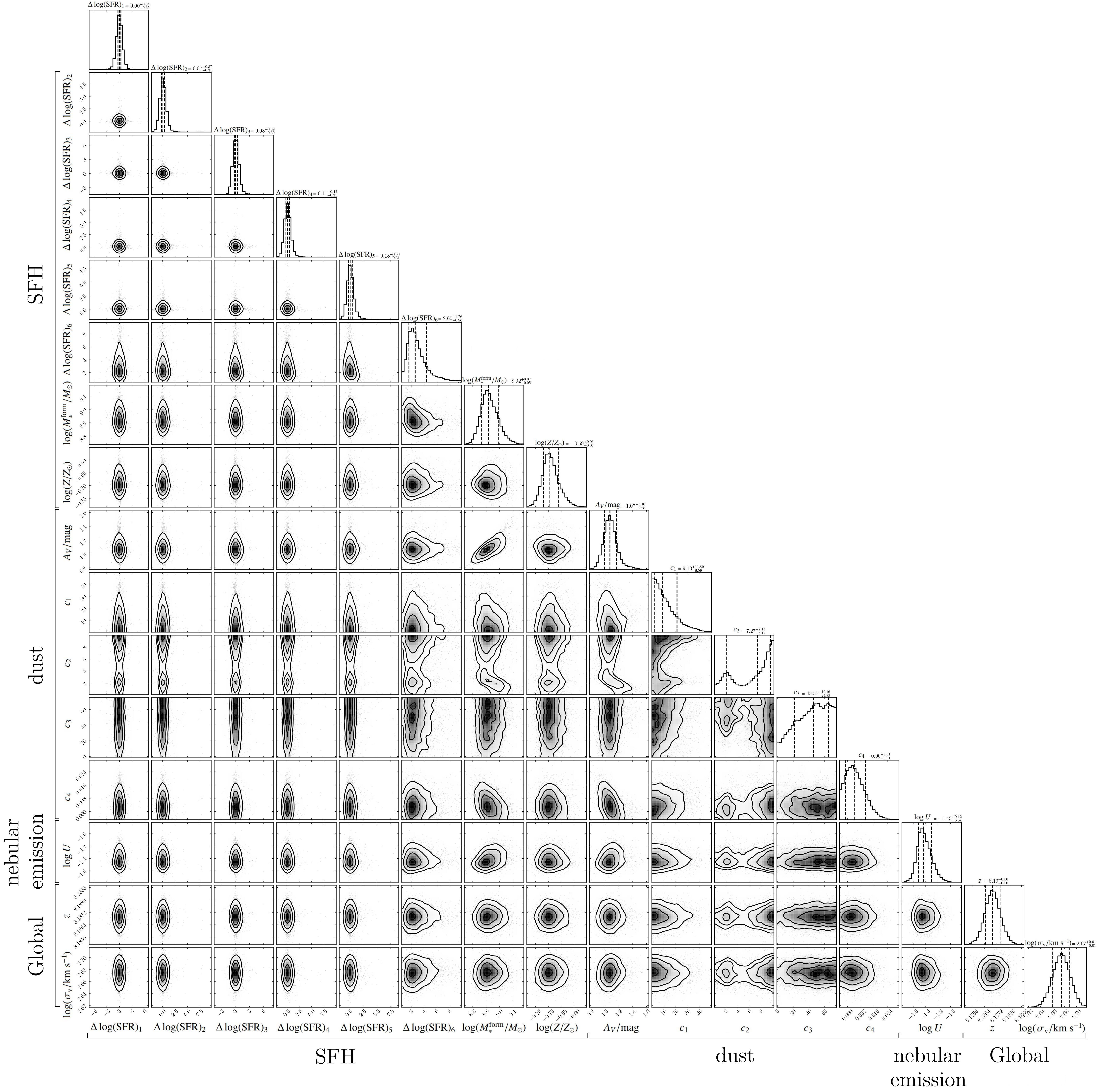}
 \caption{Corner plot illustrating the the 1D and 2D projections of the posterior distribution of the parameters derived from the SED fitting on spectrum of s01149.
     \label{corner3}
 } 
\end{figure*}

\end{appendix}
\end{document}